\documentclass[aps,twocolumn]{revtex4}
\usepackage{amsfonts}
\usepackage{amsmath}
\usepackage{amssymb,epsf}
\usepackage{color}
\usepackage{graphicx}
\usepackage{epstopdf}
\usepackage{float}
\usepackage{caption}
\usepackage{subfig}

\begin{document}

\title{Dark Energy Star in Gravity's Rainbow}
\author{A. Bagheri Tudeshki$^{1}$, G. H. Bordbar$^{1}$\footnote{
email address: ghbordbar@shirazu.ac.ir}, and B. Eslam Panah$^{2,3,4}$\footnote{
email address: eslampanah@umz.ac.ir}}

\begin{abstract}
The concept of dark energy can be a candidate for preventing the
gravitational collapse of compact objects to singularities. According to the
usefulness of gravity's rainbow in UV completion of general relativity (by
providing a new description of spacetime), it can be an excellent option to
study the behavior of compact objects near phase transition regions. In this
work, we obtain a modified Tolman-Openheimer-Volkof (TOV) equation for
anisotropic dark energy as a fluid by solving the field equations in
gravity's rainbow. Next, to compare the results with general relativity, we
use a generalized Tolman-Matese-Whitman mass function to determine the
physical quantities such as energy density, radial pressure, transverse
pressure, gravity profile, and anisotropy factor of the dark energy star. We
evaluate the junction condition and investigate the dynamical stability of
dark energy star thin shell in gravity's rainbow. We also study the energy
conditions for the interior region of this star. We show that the
coefficients of gravity's rainbow can significantly affect this non-singular
compact object and modify the model near the phase transition region.
\end{abstract}

\address{$^{1}$ Department of Physics and Biruni Observatory, Shiraz University, Shiraz 71454, Iran \\	
$^{2}$  Department of Theoretical Physics, Faculty of Science, University of Mazandaran, P. O. Box 47415-416, Babolsar, Iran\\
$^{3}$ ICRANet-Mazandaran, University of Mazandaran, P. O. Box 47415-416, Babolsar, Iran\\
$^{4}$ ICRANet, Piazza della Repubblica 10, I-65122 Pescara, Italy}

\maketitle

\section{Introduction}

The existence of a strange cosmic fluid known as dark energy (DE) with the
negative pressure, describes the universe's accelerating expansion,
requiring special consideration in other structures, including compact
objects. In these cases, it also seems that using other modified gravity is
a solution to the problem of quantum-mechanical incompatibility and general
relativity. With the link between physical concepts on the demand of the
general relativity, quantum mechanics and condensed matter physics by
Chapline \cite{Chapline}, a new era in the development of alternative
theories of black holes was formed. It may also be possible at the surface
of a black hole (event horizon) a behavior similar to that of a Bose gas. In
the sense that it is possible that the surface of a black hole is a result
of a quantum phase transition.

One of the alternative candidates to the black hole theory is gravastar. The
concept of gravastar was first introduced by Mazur and Mottola \cite%
{MazurM2004}. They introduced a compact object with three regions and
different space-times; the interior vacuum region, which is related to the
cosmological constant, the middle region, which is a shell with a finite
thickness, and the equation of state of a perfect fluid governs it, and also
the exterior region, which contains a true vacuum, and the pressure in it is
zero \cite{MazurM2004}. The advantage of this model is that there is no
singularity at the center of this compact object, and there is also no
horizon \cite{MazurM2004}. Many studies have been done on gravastar. A $5-$%
layer model similar to the introduced gravastar in Ref. \cite{MazurM2004}
plus a junction interface in the middle region was investigated by Visser
and Wiltshire \cite{VisserW2004}. For the first time, gravastars with
anisotropic pressures were studied by Cattoen et al. \cite{Cattoen2005}.
They showed that the Tolman-Oppenheimer-Volkoff (TOV) equation is not
satisfied in the shell of gravastar-type objects with isotropic pressure,
and anisotropic pressure must be considered.

The concept of the dark energy star (DES) was first introduced by Chaplin
\cite{ChaplinearXiv}. This idea holds that at a critical surface, the
falling matter is converted into vacuum energy, which is much larger than
the cosmic vacuum energy, creating a negative pressure to act against
gravity \cite{ChaplinearXiv}. Therefore, a singularity does not occur inside
the star. The observational results suggest that dark energy is very
homogeneous and not very dense. However, there is still attention to
anisotropic dark energy. Koivisto and Mota, in two works \cite%
{KoivistoI,KoivistoII} proposed a universe full of dark energy, and they
studied its features in detail. Also, in order to investigate the low
quadrupole in the CMB oscillations, an anisotropic equation of state was
attributed to dark energy \cite{Campanelli2011}. But the concept of
anisotropy in the study of compact objects is derived from the role of some
physical events such as phase transitions \cite{Sokolov1980}, a pion
condensation \cite{Hartle1975}, the presence of strong magnetic \cite%
{Bordbar2022} and electric fields \cite{Usov2004}, a solid core, etc. Lobo
and Crawford \cite{LoboC2005} generally investigated the behavior of thin
shells by benefiting from the application of the Lanczos equations and
Gauss-Kodazzi equations. Then by generalizing this method, they discussed
the stability of thin shell anent black holes and wormholes. Following this
model, the dynamical stability of DES was studied in several cases. Lobo
\cite{Lobo2006} selected two models: constant energy density and
Tolman-Matese-Whitman mass function for a dark energy star with a transverse
pressure. He showed that there are stable regions near the surface of star.
Ghezzi \cite{Ghezzii2011} proposed a model of a compact object with the
fermionic matter coupled to inhomogeneous anisotropic variable dark energy.
Then he obtained the TOV equation and physical quantities such as mass in
relation to coupling parameter. Considering the phantom scalar field as a
model of dark energy, Yazadjiev \cite{Yazadjiev2011} provided an exact
solution for the interior of the DES in the presence of matter. The
stability condition and various physical properties for a special type of
dark energy star with five regions were discussed in Ref. \cite{BharR2015}.
By introducing a mass function, Bhar et al. \cite{Bharetal2018} studied the
structure and stability of dark energy stars and compared the results with
the observational results. The time-dependent equations of motion for dark
energy stars was studied in Ref. \cite{BeltracchiG2019}. In this study, it
is assumed that the pressure of the fluid is positive at the beginning of
falling and then, where the star reaches its final stage of collapse, the
negative pressure prevails in the system. By expressing the equations of
motion in the presence of specific metric potentials Finch and Skea,
Banerjee et al. \cite{Banerjee2020} were able to provide an exact solution
for the dark energy star. The effects of slow rotation on the configuration
of a dark energy star with the governing Chaplygin equation of state were
investigated by Panotopoulos et al. in Ref. \cite{Panotopoulos2021}. In most
researches, the stability of dark energy star has been confirmed, but it has
been shown that a dark energy star can be physically unstable in the
presence of a phantom field \cite{Sakti2021}. The study of dark energy stars
at modified gravity is underway. The physical quantities of DES in the
presence of Einstein-Gauss-Bonnet gravity were determined by Malaver et al.
\cite{Malaver2021}. In other work, Bhar \cite{Bhar2021} studied physical
properties of dark energy stars such as density, pressure, mass function,
surface redshift, and maximum mass using metric potentials Tolman-Kuchowicz
(TK) and showed that all constraints are regular.

Gravitational and quantum behaviors in the phase transition layer of the
dark energy star can be a good reason to use modified gravity. In a study,
Magueijo and Smolin \cite{MagueijoS2004} by introducing rainbow functions,
suggested that the metric in a dual spacetime can depend on energy, and as a
result, the equations of motion also change. Recently, in several studies,
the effect of rainbow functions in examining different states of physical
phenomena have been investigated using the theory of gravity's rainbow \cite%
{Galan2004,Hackett2006,Aloisio2006,Ling2007,Garattini2014a,Chang2015,Santos2015}%
. Also, by attributing the energy to the location of the horizon of the two
inner and outer particles of a black hole, Ali et al. \cite{Ali2015} showed
that the information can be transferred from inside the black hole to the
outside. Thermodynamic behavior of black holes in the presence of gravity's
rainbow have been studied in Refs. \cite%
{Galan2006,LingZ2007,Ali2014,HendiPEM2016,KimKim2016,HendiFEP2016,Gangopadhyay2016,Hendi2017,
Alsaleh2017,Feng2017,EslamPanah2018,Upadhyay2018,EslamPanah2019,Morais2022,Hamil2022}%
. Energy-dependence of such geometry can produce important modifications to
non-singular compact objects \cite%
{HendiJCAP2016,Garattini2017,EslamPanah2017,Debnath2021,Mota2022}.

Our goal in this work is to study the behavior of dark energy star
properties in a modified theory of gravity is called gravity's rainbow. We
are interested in a comparison between energy-dependent physical quantities
in gravity's rainbow and energy-independent quantities in general
relativity. The plan of this paper is as follows: after introductory section
1, in section 2, we obtain field equations in gravity's rainbow, and in
section 3, junction condition and dynamic stability of thin shell are
introduced. In section 4, we determine the energy conditions, and finally, a
discussion on the results is provided in section 5.

\section{Basic Equations}

The interior spacetime ($-$) and exterior spacetime ($+$) for a spherically
symmetric metric in gravity's rainbow is given by%
\begin{equation}
ds_{\pm }^{2}=-\frac{e^{2\phi _{\pm }\left( r_{\pm }\right) }}{%
l_{\varepsilon }^{2}}dt^{2}+\frac{e^{2\lambda _{\pm }\left( r_{\pm }\right) }%
}{h_{\varepsilon }^{2}}dr^{2}+\frac{r_{\pm }^{2}\left( d\theta ^{2}+\sin
^{2}\theta d\varphi ^{2}\right) }{h_{\varepsilon }^{2}},  \label{metric}
\end{equation}%
where $e^{2\phi _{\pm }\left( r_{\pm }\right) }$ and $e^{2\lambda _{\pm
}\left( r_{\pm }\right) }$ are the metric potentials. Also, $l_{\varepsilon
}^{2}$ and $h_{\varepsilon }^{2}$ are rainbow functions. It is notable that $%
\varepsilon =E/E_{P}$, where an observer with zero acceleration measures an
energy $E$ for a test particle of mass $m$, and also $E_{P}$ refers to the
Planck energy. The modified energy-momentum dispersion is \cite%
{MagueijoS2004}%
\begin{equation}
E^{2}l_{\varepsilon }^{2}-p^{2}h_{\varepsilon }^{2}=m^{2}.
\end{equation}

The equation of motion in gravity's rainbow given by \cite{MagueijoS2004}%
\begin{equation}
G_{\mu \nu }(\varepsilon )=\frac{8\pi G(\varepsilon )}{c^{4}(\varepsilon )}%
T_{\mu \nu }\left( \varepsilon \right) ,  \label{GReq}
\end{equation}%
where $G_{\mu \nu }(\varepsilon )$ is Einstein tensor, $G(\varepsilon )$ and
$c(\varepsilon )$ are the energy-dependent gravitational constant and the
energy-dependent speed of light, respectively.\textbf{\ In the theory of
quantum gravity, a normalized gravitational coupling is defined, which
depends on energy in the scale of high energies, and as a result, }$G$%
\textbf{\ is also a function of energy. If we rewrite the linear element of Eq.
(\ref{metric}) to form }$ds^{2}=-\frac{dt^{2}}{l_{\varepsilon }^{2}}+\frac{%
\left( dx^{i}\right) ^{2}}{h_{\varepsilon }^{2}}$\textbf{, it can be seen
that the speed of light, }$c\left( \varepsilon \right) =\frac{dx}{dt}=%
\frac{h_{\varepsilon }}{l_{\varepsilon }}$\textbf{\ depends on the energy in
gravity's rainbow due to the dependence of rainbow functions on the energy. It is
necessary to note that in the limit of low energies, }$G\left( \varepsilon
\right) $\textbf{\ and }$c\left( \varepsilon \right) $\textbf{\ tend to
the universal forms }$G$\textbf{\ and }$c$\textbf{, respectively \cite%
{MagueijoS2004}.} Also $T_{\mu \nu }\left( \varepsilon \right) $ is
stress-energy tensor that plays a role as a source of time-space curvature.
Here, we assume $G(\varepsilon )=c(\varepsilon )=1$.

According to the linear element (Eq. (\ref{metric})), we assume that the
interior spacetime of DES is full of dark energy that behaves like a fluid
with equation of state $p_{r}\left( r\right) =\omega \rho \left( r\right) $,
where $\omega $ is the dark energy parameter. Note, $-1<\omega <-1/3$, $%
\omega =-1$ and $\omega <-1$ refer to the dark energy regime, the
cosmological constant and phantom energy regime, respectively. Since the
surface of a dark energy star is where the phase transition occurs, we are
interested in using dark energy as an anisotropic fluid with the transverse
and radius pressures. Generally, for an anisotropic distribution of matter,
the stress-energy tensor can be achieved as follows \cite{Bayin1986}%
\begin{eqnarray}
T_{\mu \nu }\left( \varepsilon \right)  &=&\left[ \rho \left( r\right)
+p_{t}\left( r\right) \right] u_{\mu }u_{\nu }+p_{t}\left( r\right) g_{\mu
\nu }  \notag \\
&&  \notag \\
&&+\left[ p_{r}\left( r\right) -p_{t}\left( r\right) \right] x_{\mu }x_{\nu
},  \label{T4}
\end{eqnarray}%
where $\rho \left( r\right) $, $p_{r}\left( r\right) $ and $p_{t}\left(
r\right) $ are the energy density, the radial pressure and the transverse
pressure, respectively. The transverse pressure $p_{t}\left( r\right) $ is
perpendicular to the direction of the radial pressure $p_{r}\left( r\right) $
of the fluid. Also, $u_{\mu }$ represents the four-velocity vector with $%
u_{\mu }u^{\mu }=-1$ and $x_{\mu }$ refers to the unit spacelike vector in
the radial direction which is defined by $x^{\mu }=\sqrt{\left(
g_{rr}\right) ^{-1}\delta _{~~r}^{\mu }}$\cite{Lobo2006}, and $x_{\mu
}x^{\mu }=1$. In order to obtain the modified relations in gravity's
rainbow, the metric coefficients can be converted into the following form%
\begin{eqnarray}
g_{tt} &\longrightarrow &\frac{g_{tt}}{l_{\varepsilon }^{2}},  \label{5} \\
&&  \notag \\
g_{kk} &\longrightarrow &\frac{g_{kk}}{h_{\varepsilon }^{2}},  \label{6}
\end{eqnarray}%
where $g_{tt}$ and $g_{kk}$ are metric coefficients in the general line
element, and the index $k$ refers to $r$, $\theta $ and $\varphi $. Thus by
using the interior line element Eq. (\ref{metric}) and Eqs. (\ref{T4}-\ref{6}%
), the modified components of stress-energy tensor are obtained%
\begin{eqnarray}
T_{tt}\left( \varepsilon \right)  &=&\frac{\rho \left( r\right) e^{2\phi
_{-}\left( r\right) }}{l_{\varepsilon }^{2}}, \\
&&  \notag \\
T_{rr}\left( \varepsilon \right)  &=&\frac{p_{r}\left( r\right) e^{2\lambda
_{-}\left( r\right) }}{h_{\varepsilon }^{2}}, \\
&&  \notag \\
T_{\theta \theta }\left( \varepsilon \right)  &=&\frac{p_{t}\left( r\right)
r^{2}}{h_{\varepsilon }^{2}}, \\
&&  \notag \\
T_{\varphi \varphi }\left( \varepsilon \right)  &=&\frac{p_{t}\left(
r\right) r^{2}\sin ^{2}\theta }{h_{\varepsilon }^{2}},
\end{eqnarray}%
or mixed diagonal elements of the stress-energy tensor given by%
\begin{equation}
T_{\nu }^{\mu }=diag\left[ -\rho \left( r\right) ,p_{r}\left( r\right)
,p_{t}\left( r\right) ,p_{t}\left( r\right) \right] .  \label{T}
\end{equation}
The $tt$ component of field equations (\ref{GReq}) provides the following
equation,
\begin{equation}
M_{eff}\left( r,\varepsilon \right) =\int_{0}^{r}\frac{4\pi r\prime ^{2}\rho
\left( r\prime \right) dr\prime }{h_{\varepsilon }^{2}}=\frac{m\left(
r\right) }{h_{\varepsilon }^{2}},
\end{equation}%
where $M_{eff}\left( r,\varepsilon \right) $ is the effective mass and $%
m\left( r\right) $ refers the mass function. By calculating the $rr$
component of field equations (\ref{GReq}), we get the following relation%
\begin{equation}
\frac{d\phi _{-}}{dr}=g\left( r\right) =\frac{M_{eff}\left( r,\varepsilon
\right) h_{\varepsilon }^{2}+4\pi r^{3}p_{r}\left( r\right) }{r\left(
r-2M_{eff}\left( r,\varepsilon \right) \right) h_{\varepsilon }^{2}},
\label{phii}
\end{equation}%
where $g\left( r\right) $ is called "gravity profile" \cite{Lobo2006} which
is related to the local acceleration in gravity's rainbow which is
represented by $A=\sqrt{h_{\varepsilon }^{2}e^{-2\lambda \left( r\right) }}%
g\left( r\right) $, and it is related to the redshift function by $\phi
\left( r\right) =-\int_{r}^{\infty }g\left( \widetilde{r}\right) d\widetilde{%
r}$ \cite{Lobo2006,Morris1988}. If $g\left( r\right) >0$, local acceleration
due to gravity of the interior solution be attractive and if $g\left(
r\right) <0$, local acceleration be repulsive. According to the dark energy
equation of state, we can rewrite the gravity profile in relation to the
dark energy parameter%
\begin{equation}
g\left( r\right) =\frac{M_{eff}\left( r,\varepsilon \right) +r\omega \left(
\frac{\partial M_{eff}\left( r,\varepsilon \right) }{\partial r}\right) }{%
r\left( r-2M_{eff}\left( r,\varepsilon \right) \right) }.  \label{g(r)I}
\end{equation}
Using conservation law $\triangledown ^{\mu }T_{\mu \nu }=0$, and inserting
Eq. (\ref{phii}) into it, we obtain the TOV equation for an anisotropic
distribution of matter in gravity's rainbow%
\begin{eqnarray}
\frac{dp_{r}\left( r\right) }{dr} &=&-\frac{\left( 4\pi r^{3}p_{r}\left(
r\right) +h_{\varepsilon }^{2}M_{eff}\left( r,\varepsilon \right) \right) %
\left[ p_{r}\left( r\right) +\rho \left( r\right) \right] }{r\left(
r-2M_{eff}\left( r,\varepsilon \right) \right) h_{\varepsilon }^{2}}  \notag
\\
&&  \notag \\
&&+\frac{2\left[ p_{t}\left( r\right) -p_{r}\left( r\right) \right] }{r}.
\label{ModTOVI}
\end{eqnarray}
We can define the anisotropy factor $\Delta \left( r\right) =p_{t}\left(
r\right) -p_{r}\left( r\right) $, and write both sides of the above equation
in terms of $M_{eff}\left( r,\varepsilon \right) $ and dark energy
parameter, hence $\Delta \left( r\right) $ is written as follows%
\begin{eqnarray}
\Delta \left( r\right)  &=&\frac{\omega h_{\varepsilon }^{2}}{8\pi r^{2}}%
\left[ r\left( \frac{\partial ^{2}M_{eff}\left( r,\varepsilon \right) }{%
\partial r^{2}}\right) -2\left( \frac{\partial M_{eff}\left( r,\varepsilon
\right) }{\partial r}\right) \right.   \notag \\
&&  \notag \\
&&\left. +\left( \frac{\left( 1+\omega \right) r}{\omega }\right) r\left(
\frac{\partial M_{eff}\left( r,\varepsilon \right) }{\partial r}\right)
g\left( r\right) \right] .  \label{Delta}
\end{eqnarray}%
Also, $\frac{\Delta \left( r\right) }{r}$ indicates a force caused by the
anisotropic behaviors of the stellar model. If $\Delta \left( r\right) >0$,
this force is repulsive, but if $\Delta \left( r\right) <0$, this force is
attractive. In order to have a standard solution for the dark energy stars, $%
\Delta \left( r\right) $ should be positive. Both the gravity profile $%
g\left( r\right) $\ and the anisotropy factor $\Delta \left( r\right) $
depend on $h_{\varepsilon }^{2}$.

To solve the field equations, we have to guess a suitable mass function. To
compare the results of two different gravitational models, the general
relativity and gravity's rainbow with the same mass function model, let us
use the Tolman-Matese-Whitman (TMW) mass function that Lobo had previously
used in his study \cite{Lobo2006}. Thus we consider a modified TMW mass
function for gravity's rainbow as%
\begin{equation}
M_{eff}\left( r,\varepsilon \right) =\frac{b_{0}r^{3}}{2\left(
1+2b_{0}r^{2}\right) h_{\varepsilon }^{2}},  \label{MII}
\end{equation}%
where $b_{0}$ is a positive constant \cite{Lobo2006}, and this mass function
is regular at the origin as $r\longrightarrow 0$. We use Eqs. (\ref{g(r)I})-%
(\ref{MII}) to calculate the physical quantities of the dark energy star in
gravity's rainbow, which are%
\begin{eqnarray}
\rho \left( r\right)  &=&\frac{b_{0}\left( 2b_{0}r^{2}+3\right) }{8\pi
\left( 2b_{0}r^{2}+1\right) },  \label{rhoo} \\
&&  \notag \\
p_{r}\left( r\right)  &=&\frac{\omega b_{0}\left( 2b_{0}r^{2}+3\right) }{%
8\pi \left( 2b_{0}r^{2}+1\right) },  \label{Pp} \\
&&  \notag \\
g\left( r\right)  &=&\frac{2\omega b_{0}r^{2}+2b_{0}r^{2}+3\omega +1}{%
2r\left( 2b_{0}r^{2}+1\right) \left( \frac{h_{\varepsilon }^{2}\left(
2b_{0}r^{2}+1\right) }{b_{0}r^{2}}-1\right) }, \\
&&  \notag \\
\Delta \left( r\right)  &=&\frac{\omega h_{\varepsilon }^{2}\mathcal{A}_{1}-%
\frac{b_{0}^{2}r^{4}\left( \mathcal{A}_{2}+\frac{\mathcal{A}_{3}}{b_{0}r^{2}}%
\right) }{2}-\frac{3\mathcal{A}_{4}}{8}}{\frac{4\pi \left(
2b_{0}r^{2}+1\right) ^{3}}{-b_{0}}\left( \frac{h_{\varepsilon }^{2}\left(
2b_{0}r^{2}+1\right) }{b_{0}r^{2}}-1\right) },  \label{Deltaa}
\end{eqnarray}%
where $\mathcal{A}_{1}$, $\mathcal{A}_{2}$, $\mathcal{A}_{3}$ and $\mathcal{A%
}_{4}$ are%
\begin{eqnarray*}
\mathcal{A}_{1} &=&4b_{0}^{2}r^{4}+12b_{0}r^{2}+5, \\
&& \\
\mathcal{A}_{2} &=&\omega ^{2}+6\omega +1, \\
&& \\
\mathcal{A}_{3} &=&3\omega ^{2}+15\omega +2, \\
&& \\
\mathcal{A}_{4} &=&\omega ^{2}+4\omega +1.
\end{eqnarray*}
Note that using the energy density relation Eq. (\ref{rhoo}), and defining
the central energy density $\rho _{c}$ in $r=0$, the constant $b_{0}$ is
obtained $b_{0}=8\pi \rho _{c}/3$.

\begin{figure}[tbh]
\centering
\includegraphics[width=0.22\textwidth]{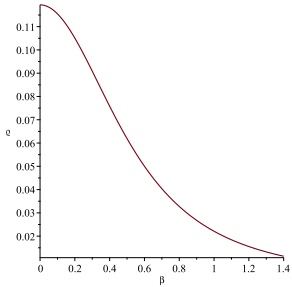} \newline
\caption{Dimensionless energy density ($\protect\varrho =\frac{\protect\rho %
\left( r\right) }{b_{0}}$) versus dimensionless parameter ($\protect\beta =%
\protect\sqrt{b_{0}}r$) with modified TMW mass function.}
\label{Fig1}
\end{figure}

\begin{figure}[tbh]
\centering
\includegraphics[width=0.22\textwidth]{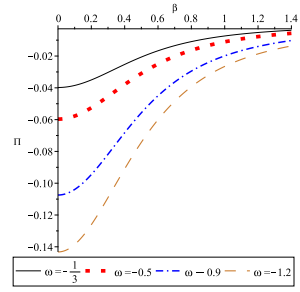} \newline
\caption{Dimensionless radial pressure ($\Pi =\frac{p_{r}\left( r\right) }{%
b_{0}}$) versus dimensionless parameter ($\protect\beta =\protect\sqrt{b_{0}}%
r$) with modified TMW mass function for different values of $\protect\omega $%
.}
\label{Fig2}
\end{figure}

Figs. \ref{Fig1} and \ref{Fig2} show the behavior of energy density $%
\rho \left( r\right) $ and radial pressure $p_{r}\left( r\right) $ relative
to the distance from the center of star, respectively. Note that in order to
make the distance as a dimensionless quantity, the parameter $\beta $ is
defined by $\beta =\sqrt{b_{0}}r$. $\rho \left( r\right) $ and $p_{r}\left(
r\right) $ are independent of the rainbow function. Fig. \ref{Fig2}
illustrates that as the value $\omega $ of increases, the magnitude of the
radial pressure increases. The negative radial pressure is one of the
characteristics of dark energy.
%

\begin{figure}[tbh]
\centering
\includegraphics[width=0.23\textwidth]{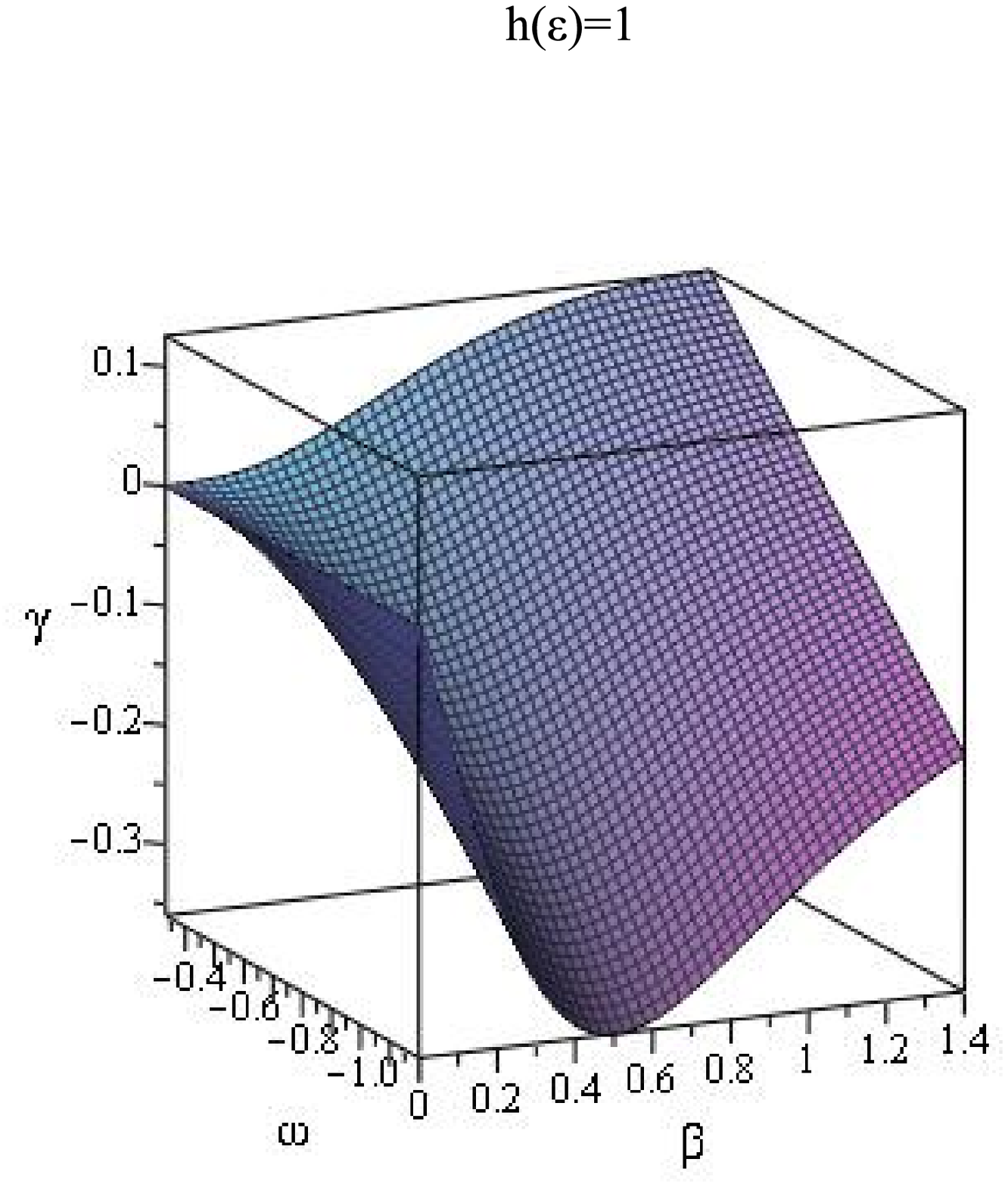} %
\includegraphics[width=0.23\textwidth]{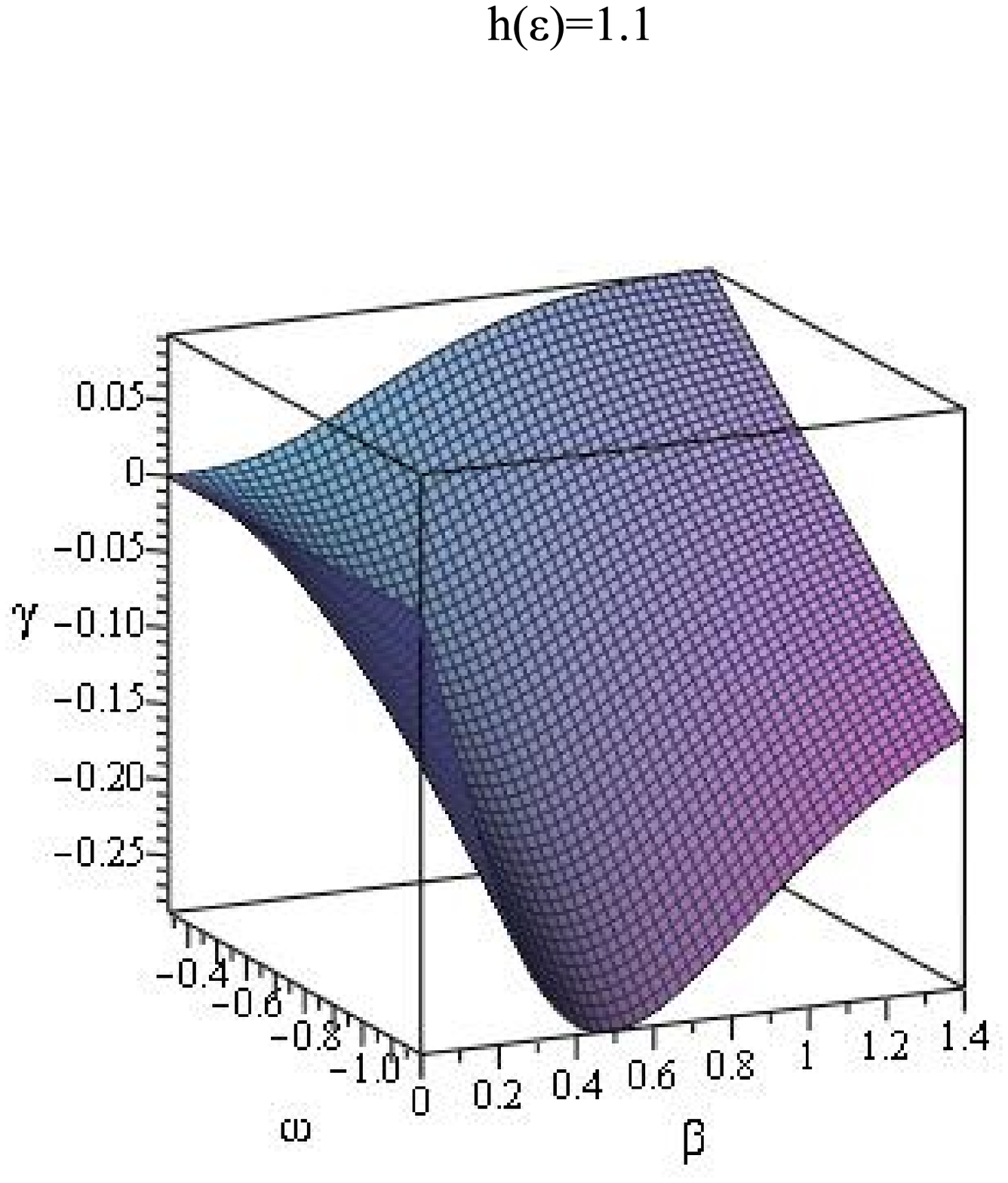} \newline
\includegraphics[width=0.23\textwidth]{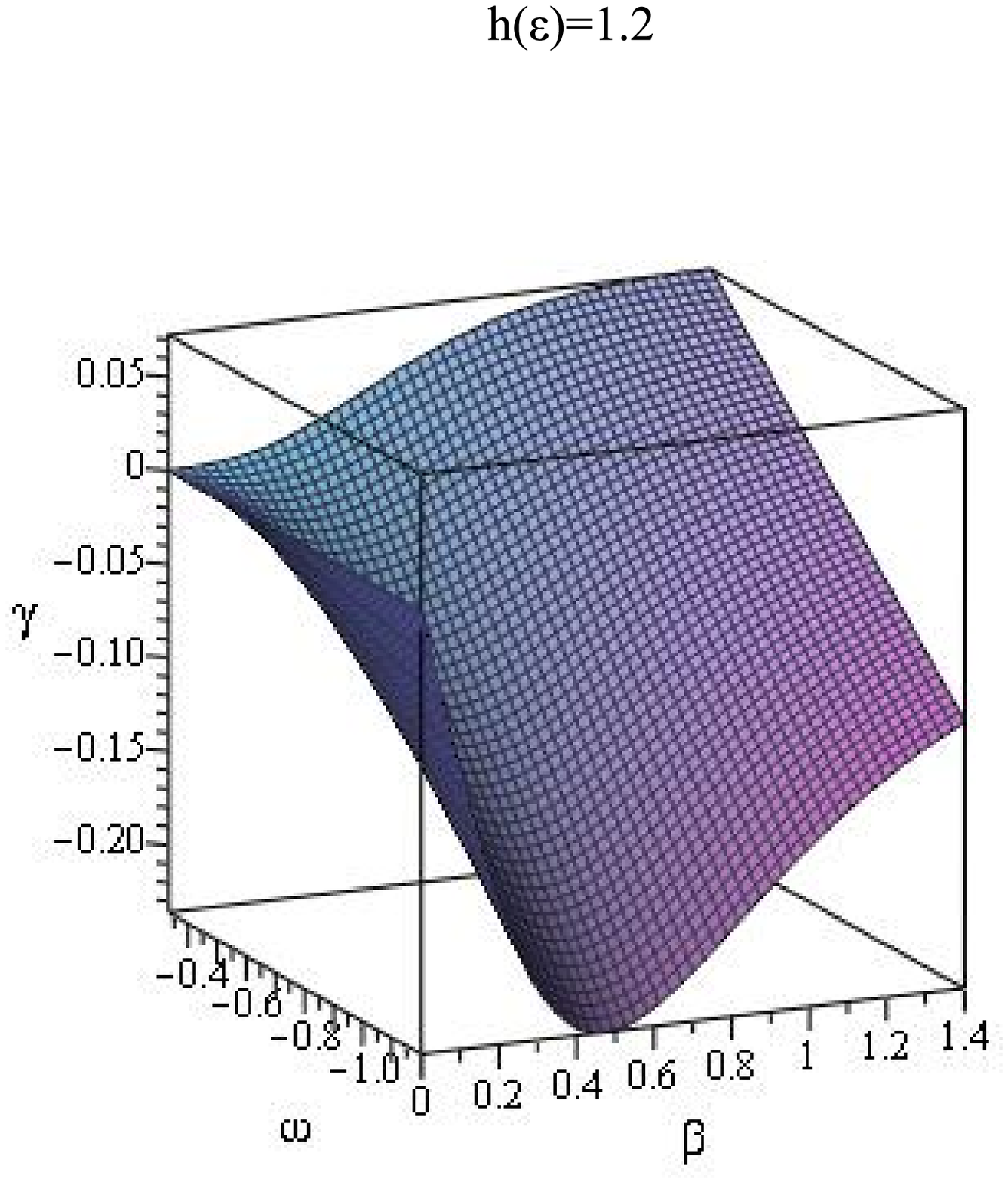} %
\includegraphics[width=0.23\textwidth]{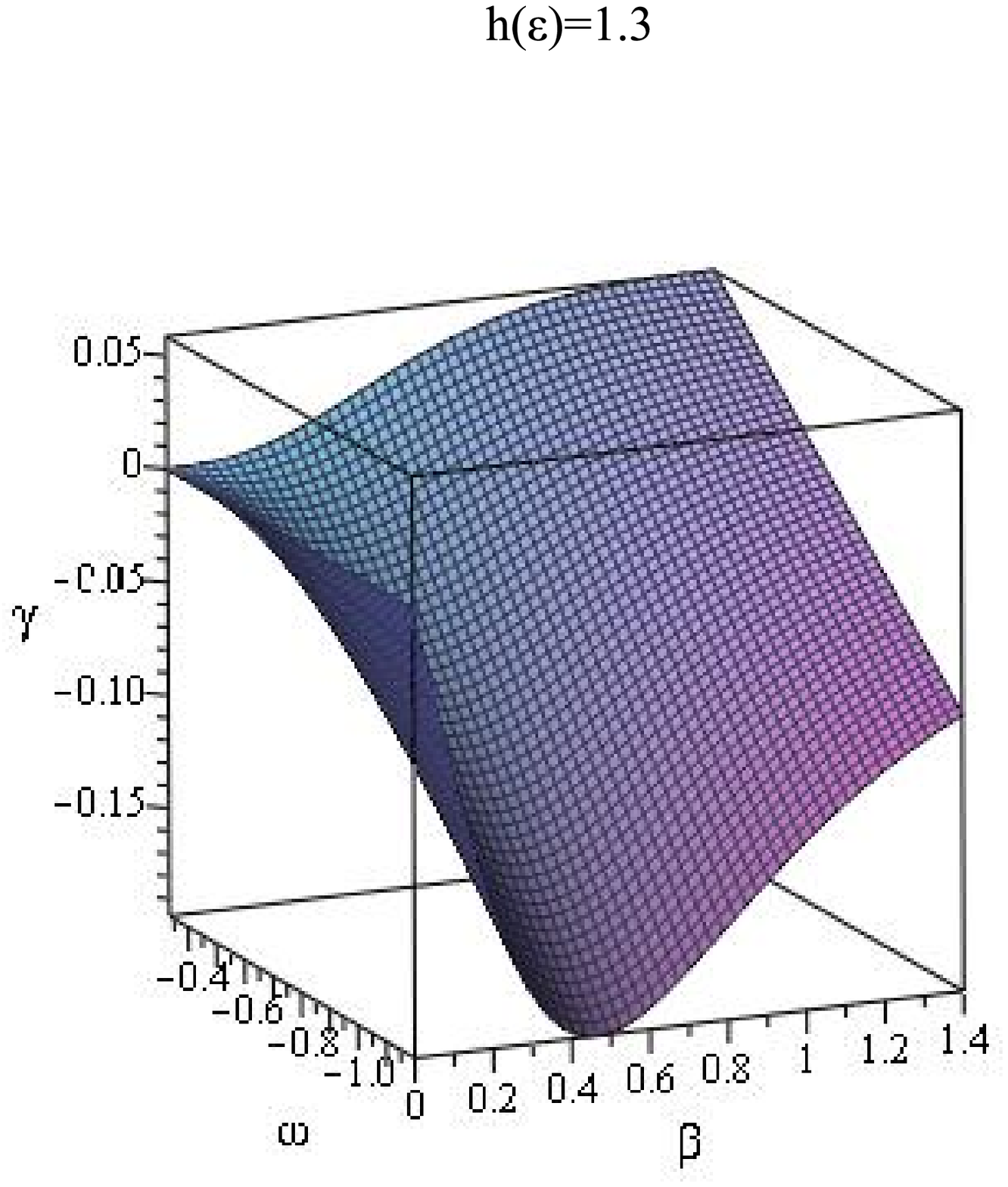} \newline
\caption{Three-dimensional diagram of the dimensionless gravity profile ($%
\protect\gamma =\frac{g(r) }{b_{0}}$) versus dimensionless parameter ($%
\protect\beta =\protect\sqrt{b_{0}}r$) with modified TMW mass function for
different values of $h_{\varepsilon }$. $h_{\varepsilon }=1$ (up left panel), $h_{\varepsilon }=1.1$ (up right panel), $h_{\varepsilon }=1.2$
(down left panel), and $h_{\varepsilon }=1.3$ (down right
panel).}
\label{Fig3}
\end{figure}

To maintain the gravitational stability of DES, $g\left( r\right) $ should
be negative. The gravity profile behavior is plotted versus $\omega $ and $%
\beta $ in both dark energy and phantom energy regimes for different values
of $h_{\varepsilon }$ in Fig. \ref{Fig3}. The range of $\beta $ is
numerically determined according to the standard $g\left( r\right) $ range
and $\omega $ values. Gravity profile values in the vicinity $\omega =-1/3$
are positive. As $h_{\varepsilon }$ increases, the range of $g\left(
r\right) $ becomes more constrained and its positive values decrease. By
reducing or removing the positive values of gravity profile, the model gets
closer to the standard model of the dark energy star. Note that $%
h_{\varepsilon }=1$, refers to the gravity profile in general relativity
\cite{Lobo2006}.

\begin{figure}[tbh]
\centering
\includegraphics[width=0.23\textwidth]{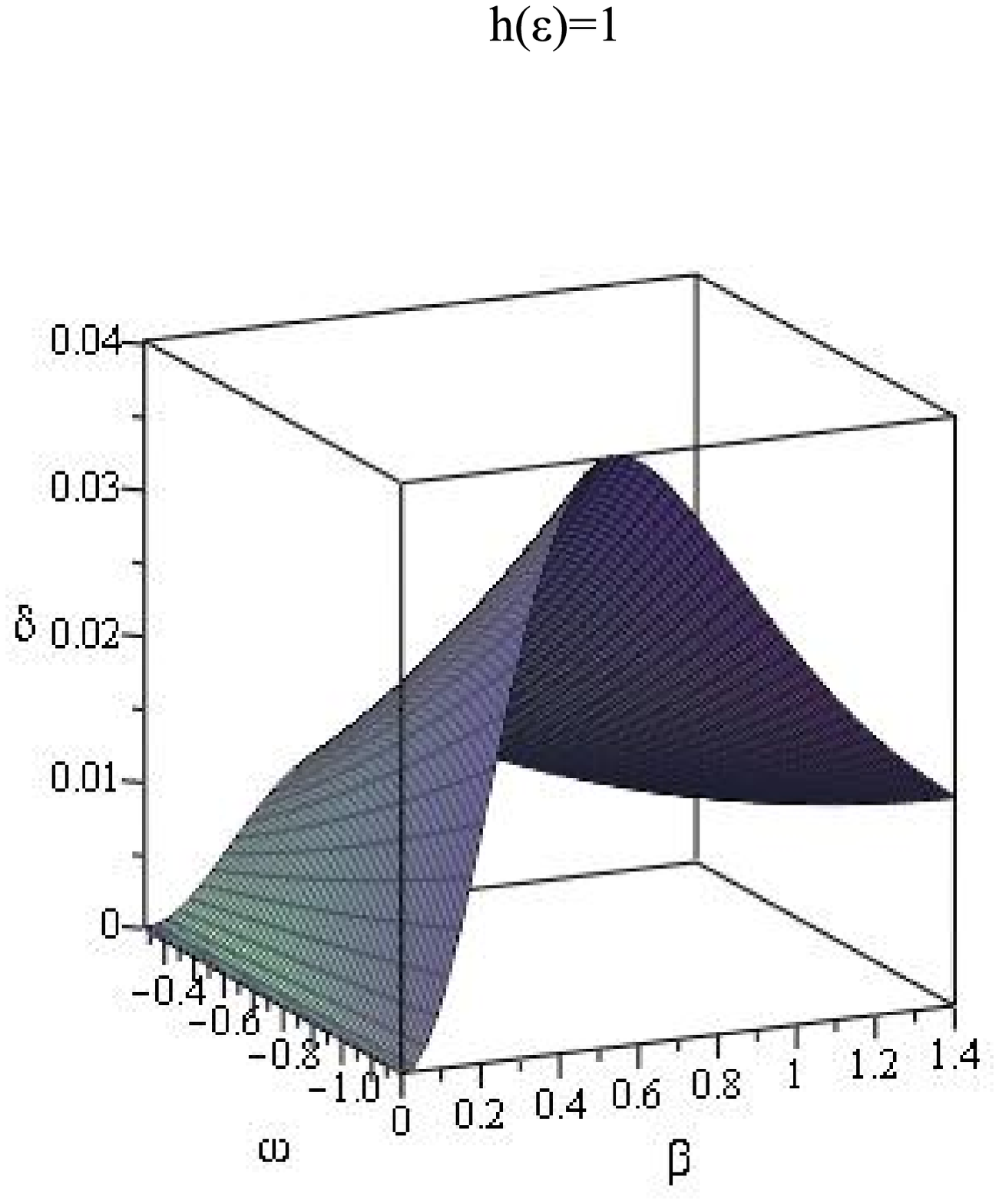} %
\includegraphics[width=0.23\textwidth]{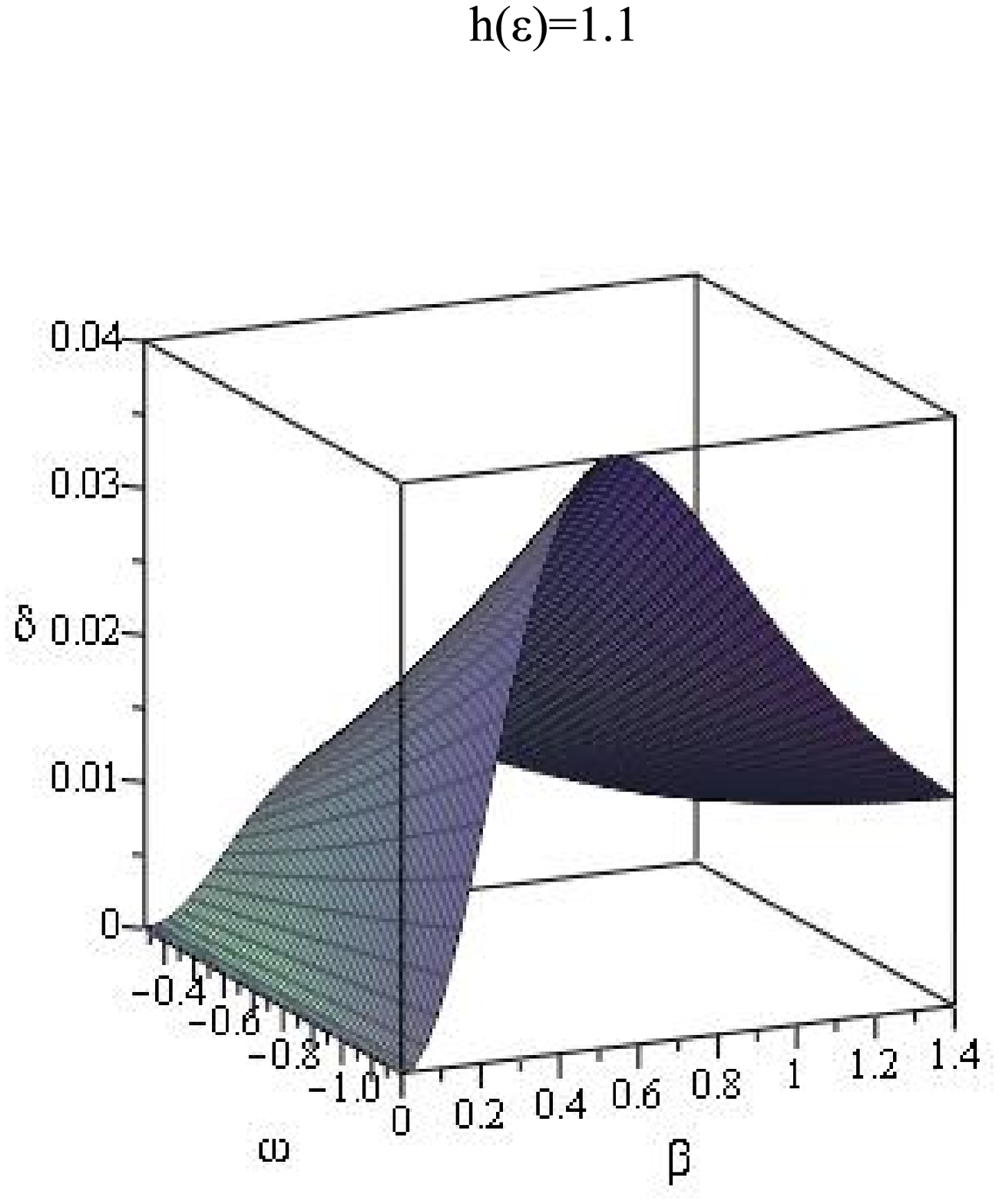} \newline
\includegraphics[width=0.23\textwidth]{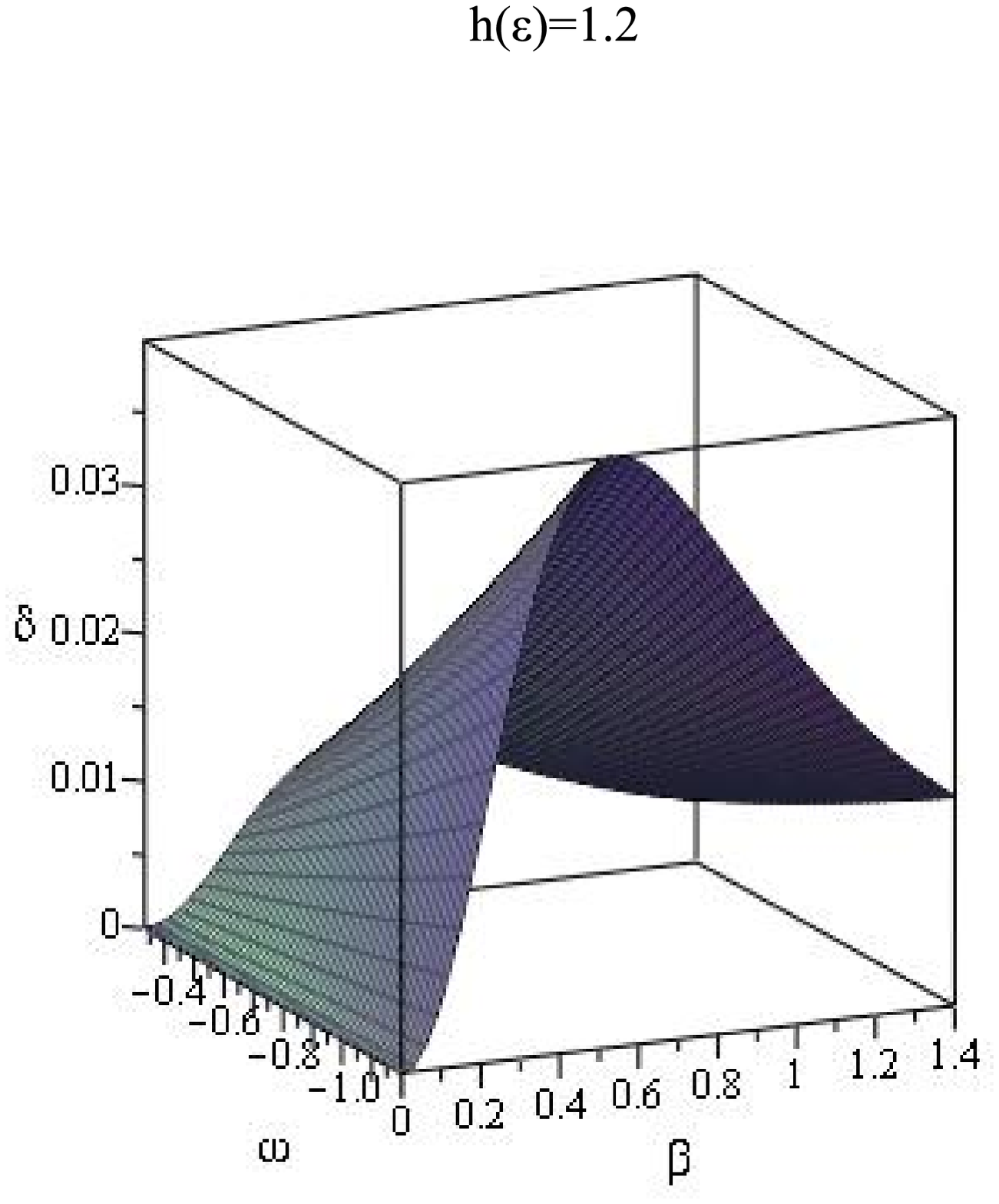} %
\includegraphics[width=0.23\textwidth]{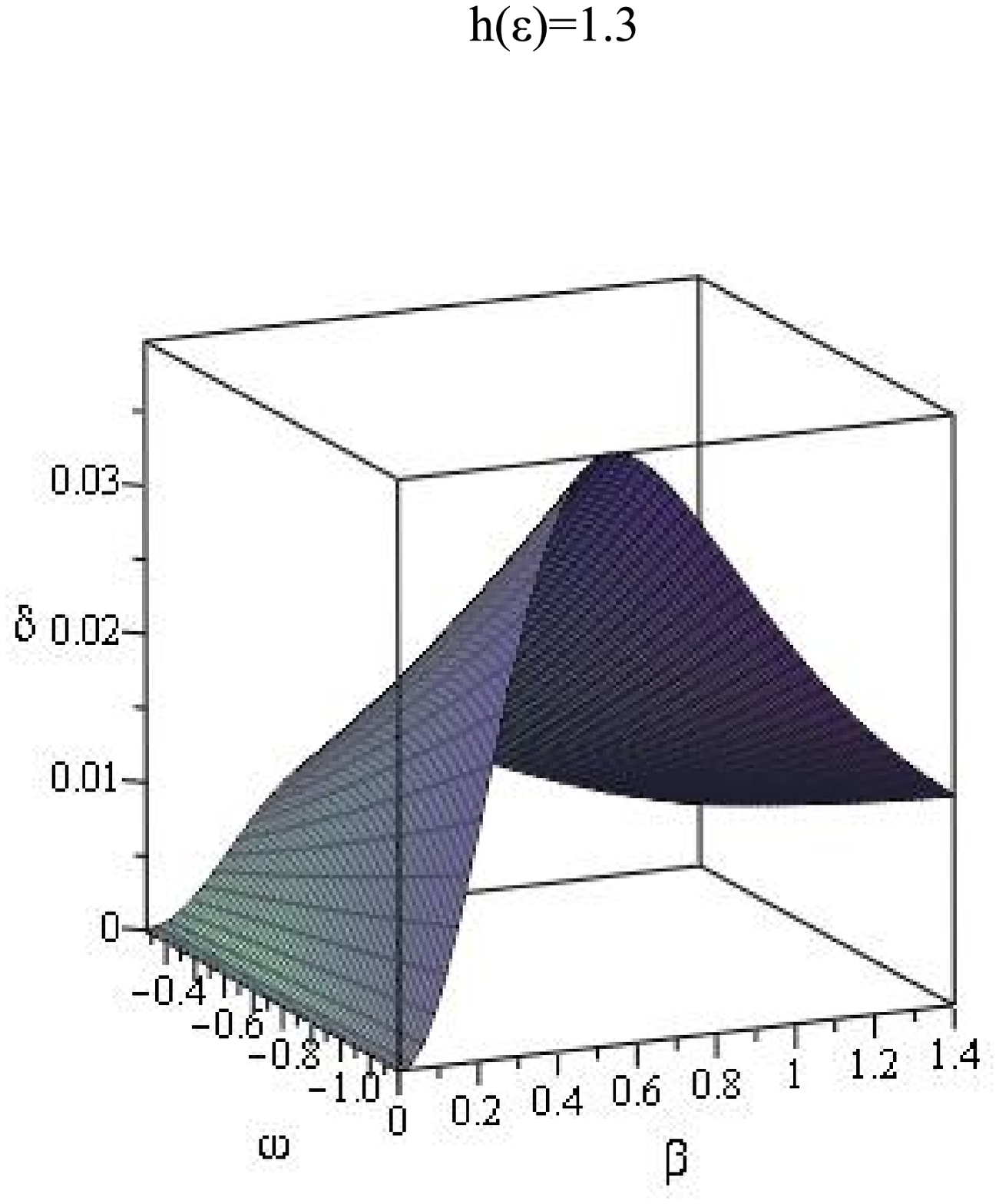} \newline
\caption{Three-dimensional diagram of the dimensionless anisotropy factor ($%
\protect\delta =\frac{\Delta }{b_{0}}$) versus dimensionless parameter ($%
\protect\beta =\protect\sqrt{b_{0}}r$) with modified TMW mass function for
different values of $h_{\varepsilon }$. $h_{\varepsilon }=1$ (up left panel), $h_{\varepsilon }=1.1$ (up right panel), $h_{\varepsilon }=1.2$
(down left panel), and $h_{\varepsilon }=1.3$ (down
right panel).}
\label{Fig4}
\end{figure}

\begin{figure}[tbh]
\centering
\includegraphics[width=0.22\textwidth]{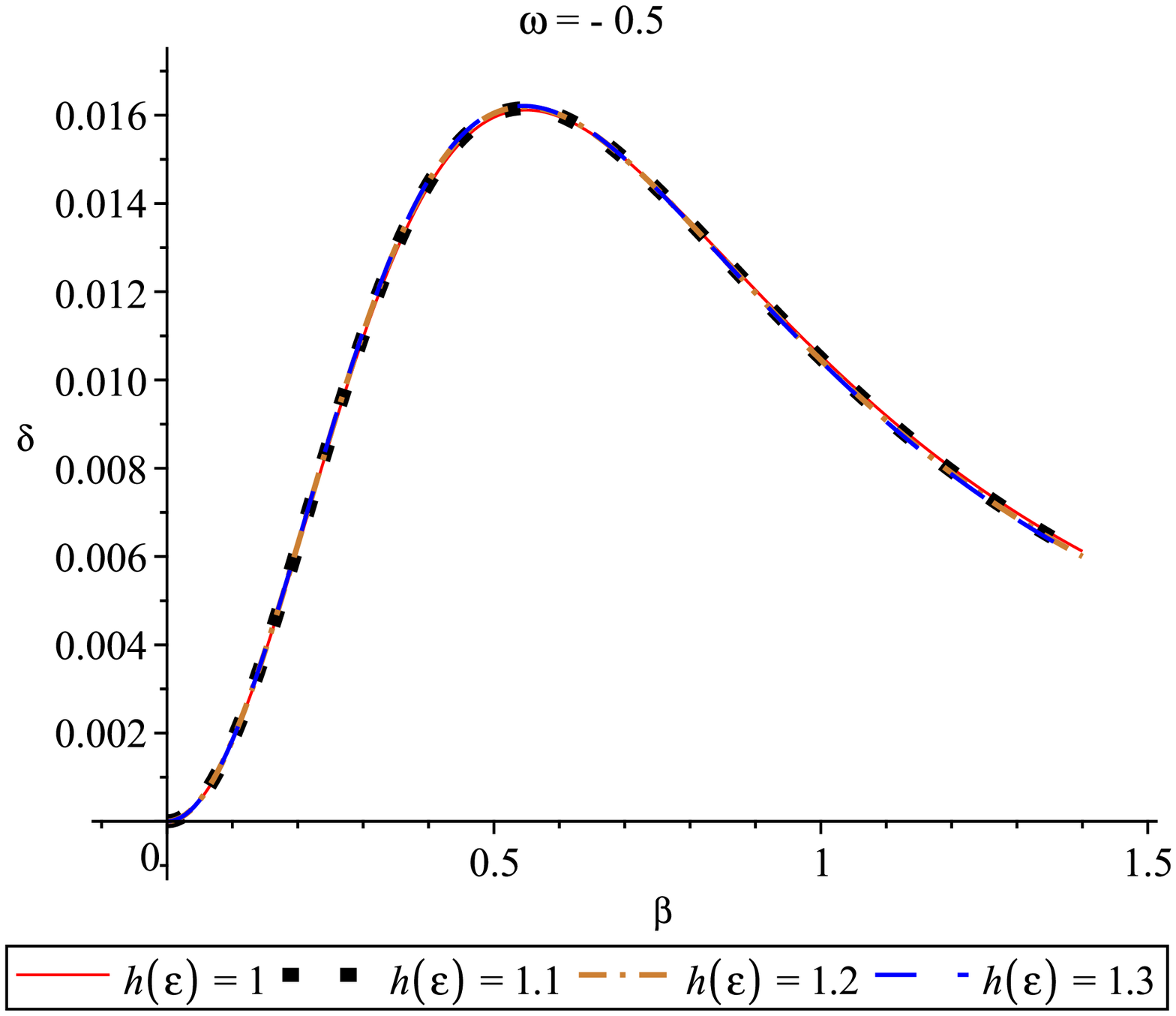} %
\includegraphics[width=0.22\textwidth]{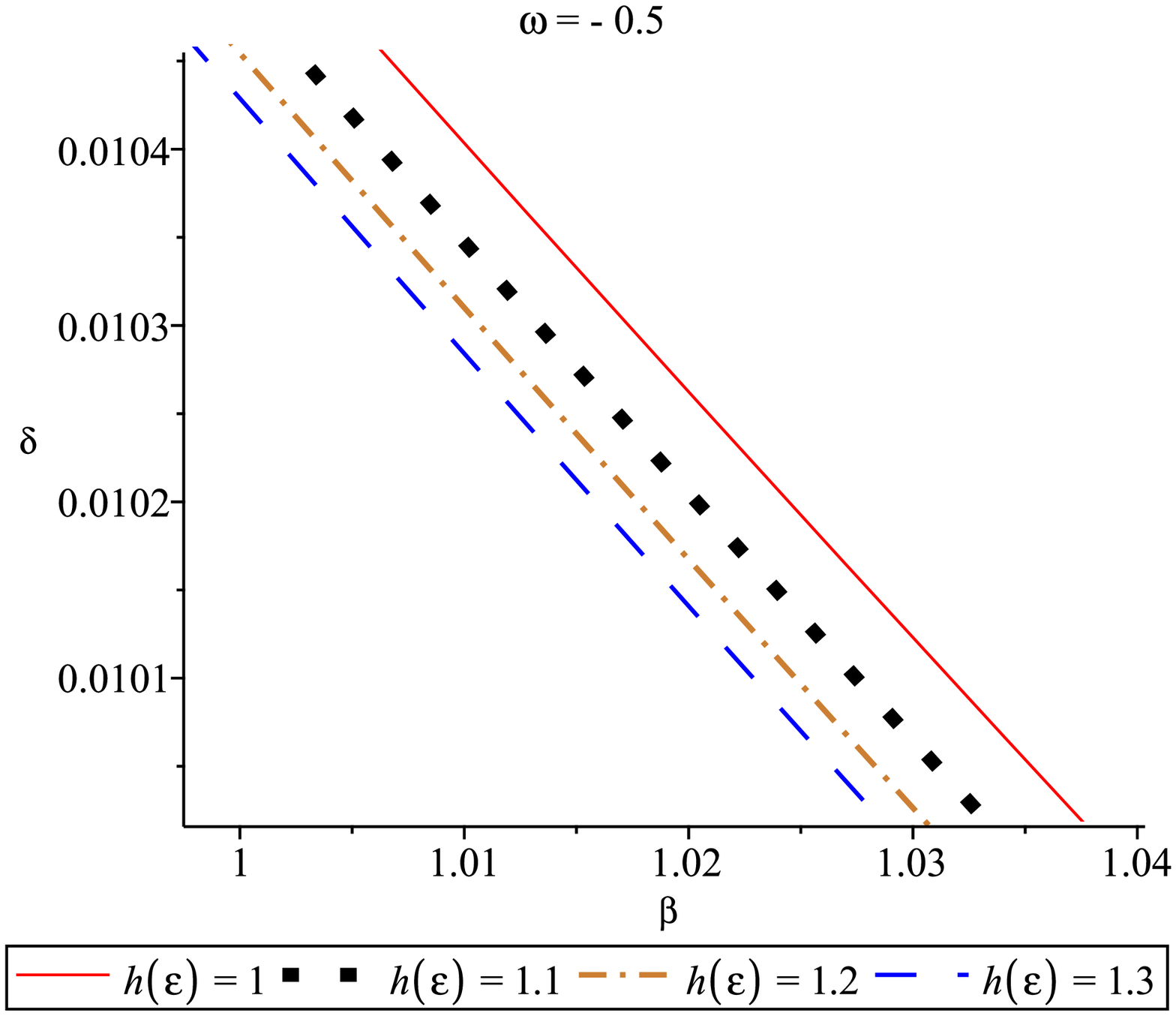} \newline
\caption{Dimensionless anisotropy factor ($\protect\delta=\frac{\Delta(r) }{%
b_{0}}$) versus dimensionless parameter ($\protect\beta =\protect\sqrt{b_{0}}%
r$) with modified TMW mass function for different values of $h_{\varepsilon }$.}
\label{Fig5}
\end{figure}

The anisotropy factor is shown in Figs. \ref{Fig4} and \ref{Fig5}. It is
observed that the anisotropy factor is positive for all $\omega $\ values.
There is also a slight difference between the anisotropy factor scheme with $%
h_{\varepsilon }=1$ (general relativity) and $h_{\varepsilon }=1.1$,$~1.2$,
and $1.3$ (gravity's rainbow).

\section{Junction Condition and Dynamic Stability of Thin Shell}

\subsection{General Relativity}

According to Darmois-Israel formalism in general relativity \cite{Israel1966}%
, we visualize two manifolds $M_{+}$ and $M_{-}$ with metrics $g_{\mu \nu
}^{\pm }\left( x_{\pm }^{\mu }\right) $. These are matched together by two
hypersurfaces $\Sigma _{\pm }$ with induced metrics $g_{ij}^{\pm }\left( \xi
\right) $, where $\xi $ is the intrinsic coordinate of hypersurfaces. Note $%
\mu ,\nu =0,1,2,3$ refer the coordinates of the $4-$dimensional manifold,
and $i,j=1,2,3$ refer the coordinates of the $3-$dimensional shell. The
induced metric on the junction surface is defined by the following relation
\cite{Israel1966}%
\begin{equation}
g_{ij}=\left[ g_{\mu \nu }\frac{\partial x^{\mu }}{\partial \xi ^{i}}\frac{%
\partial x^{\nu }}{\partial \xi ^{j}}\right] _{\pm }.  \label{imetric}
\end{equation}
We select the parametric equation for a timelike hypersurface $\Sigma $ in
the form $f\left( r,\tau \right) =r-a\left( \tau \right) =0$. The junction
radius $a\left( \tau \right) $ is a function of proper time $\tau $. It is
notable that $ds^{2}$ must be continuous throughout the junction. Using Eq. (%
\ref{imetric}), the intrinsic metric to $\Sigma $ is written by \cite%
{Lobo2004}%
\begin{equation}
ds_{\Sigma }^{2}=-d\tau ^{2}+a^{2}\left( \tau \right) \left( d\theta
^{2}+\sin ^{2}\theta d\varphi ^{2}\right) .
\end{equation}
According to the parametric equation, it can be shown that the unit normal
to the junction surfaceare $n_{\mu }$ defined as follows \cite{Poisson2004}%
\begin{equation}
n_{\mu }=\pm \frac{\frac{\partial f}{\partial x^{\mu }}}{\sqrt{g^{\alpha
\beta }\frac{\partial f}{\partial x^{\alpha }}\frac{\partial f}{\partial
x^{\beta }}}},  \label{etaI}
\end{equation}%
where $n^{\mu }n_{\mu }=+1$, and $u^{\mu }n_{\mu }=0$. Let us use the
extrinsic curvature tensor $\kappa _{ij}$ of junction surface \cite%
{LoboC2005}%
\begin{equation}
\kappa _{ij}=-n_{\mu }\left( \frac{\partial ^{2}x^{\mu }}{\partial \xi
^{i}\partial \xi ^{j}}+\Gamma _{\alpha \beta }^{\mu \pm }\frac{\partial
x^{\alpha }}{\partial \xi ^{i}}\frac{\partial x^{\beta }}{\partial \xi ^{j}}%
\right) .  \label{kappaI}
\end{equation}
The cause of the discontinuity in the extrinsic curvature is the presence of
matter in the shell \cite{Mansouri1996}, thus the discontinuity in the
extrinsic curvature is defined as \cite{Lobo2004}%
\begin{equation}
\chi _{ij}=\kappa _{ij}^{+}-\kappa _{ij}^{-},
\end{equation}%
and we can define the surface stress-energy tensor on $\Sigma $ \cite%
{Visser1989,Poisson1995}%
\begin{equation}
s_{\ j}^{i}=\frac{-1}{8\pi }\left( \chi _{~j}^{i}-\delta _{~j}^{i}\chi
_{~k}^{k}\right) .  \label{sform}
\end{equation}
This relation is known as the Lanczos equation, which roughly shows the
dynamic behavior of the thin shell. We can obtain the non-zero components of
the extrinsic curvature tensor $\kappa _{ij}$ \cite{LoboC2005}, by using
Eqs. (\ref{etaI}) and (\ref{kappaI})%
\begin{eqnarray}
\kappa _{\theta }^{\theta \pm } &=&\frac{\sqrt{e^{-2\lambda _{\pm }\left(
r\right) }+\overset{.}{a}^{2}}}{a},  \label{kappa2} \\
&&  \notag \\
\kappa _{\tau }^{\tau \pm } &=&\frac{\phi _{\pm }^{^{\prime }}\left(
e^{-2\lambda _{\pm }\left( r\right) }+\overset{.}{a}^{2}\right) +\ddot{a}+%
\overset{.}{a}^{2}\lambda _{\pm }^{^{\prime }}}{\sqrt{e^{-2\lambda _{\pm
}\left( r\right) }+\overset{.}{a}^{2}}}.  \label{kappa3}
\end{eqnarray}
\textbf{Here, we consider }$s_{\ j}^{i}=diag\left( -\sigma ,P,P\right) $\textbf{\
where }$\sigma $\textbf{\ is the surface energy density and }$P$\textbf{\ is
tangential surface pressure.} Also, the prime denotes a derivative with
respect to junction radius "$a$" and the overdot denotes a derivative with
respect to the proper time, $\tau $. By using the Lanczos equation and Eqs. (%
\ref{kappa2}) and (\ref{kappa3}), the surface energy density and the surface
pressure can be written as follow%
\begin{eqnarray}
\sigma  &=&-\frac{\chi _{\theta }^{\theta }}{4\pi }=\frac{\sqrt{e^{-2\lambda
_{-}\left( r\right) }+\overset{.}{a}^{2}}-\sqrt{e^{-2\lambda _{+}\left(
r\right) }+\overset{.}{a}^{2}}}{4\pi a}, \\
&&  \notag \\
P &=&\frac{\chi _{\tau }^{\tau }+\chi _{\theta }^{\theta }}{8\pi }  \notag \\
&&  \notag \\
&=&\frac{\left[ \frac{\left( 1+\phi ^{^{\prime }}a\right) \left(
e^{-2\lambda \left( r\right) }+\overset{.}{a}^{2}\right) +a\ddot{a}+\lambda
^{^{\prime }}a\overset{.}{a}^{2}}{\sqrt{e^{-2\lambda \left( r\right) }+%
\overset{.}{a}^{2}}}\right] ^{\pm }}{8\pi a},
\end{eqnarray}%
where contractually, $\left[ X\right] ^{\pm }=X^{+}\left\vert \Sigma \right.
-X^{-}\left\vert \Sigma \right. $ is displayed. Poisson and Visser \cite%
{Poisson1995} defined the $\eta =\frac{\sigma ^{^{\prime }}}{P^{^{\prime }}}$
parameter, that $\sqrt{\eta }$ is the speed of sound. In the surface layer,
it should be in range $0<\eta \leq 1$. By determining $\eta $, the stability
regions can be identified.

\subsection{Gravity's Rainbow}

To study thin shell and junction conditions in gravity's rainbow, we can
define the intrinsic metric to $\Sigma $ in gravity's rainbow Eq. (\ref%
{imetric}) and it given by \cite{Amirabi2018}%
\begin{equation}
ds_{\Sigma (\text{rainbow})}^{2}=-d\tau ^{2}+\frac{a^{2}\left( \tau \right)
}{h_{\varepsilon }^{2}}\left( d\theta ^{2}+\sin ^{2}\theta d\varphi
^{2}\right),
\end{equation}%
\textbf{where }$\left( \tau ,\theta ,\varphi \right) $\textbf{\ refer the
intrinsic coordinates}. The line element should be continuous across $\Sigma $%
, therefore $\overset{.}{t}=\frac{\partial t}{\partial \tau }$ is given by%
\begin{equation}
\overset{.}{t}_{(\text{rainbow})}=l_{\varepsilon }e^{\left( \lambda -\phi
\right) _{\pm }}\sqrt{e^{-2\lambda _{\pm }\left( r\right) }+\frac{\overset{.}%
{a}^{2}}{h_{\varepsilon }^{2}}}.
\end{equation}
\textbf{The position of thin shell is given by }$x^{\mu }=\left( t\left(
\tau \right) ,a\left( \tau \right) ,\theta ,\varphi \right) $\textbf{, thus
the }$4-$\textbf{velocity can be written as}%
\begin{equation}
u_{\pm (\text{rainbow})}^{\mu }=\left( l_{\varepsilon }e^{\left( \lambda
-\phi \right) _{\pm }}\sqrt{e^{-2\lambda _{\pm }\left( r\right) }+\frac{%
\overset{.}{a}^{2}}{h_{\varepsilon }^{2}}},\overset{.}{a},0,0\right) .
\end{equation}
By using Eqs. (\ref{metric}) and (\ref{etaI}), we obtain%
\begin{equation}
n_{\pm (\text{rainbow})}^{\mu }=\left( \frac{l_{\varepsilon }\overset{.}{a}%
e^{\left( \lambda -\phi \right) _{\pm }}}{h_{\varepsilon }},\sqrt{%
h_{\varepsilon }^{2}e^{-2\lambda _{\pm }\left( r\right) }+\overset{.}{a}^{2}}%
,0,0\right) .
\end{equation}
General form of $G_{\mu \nu }\left( \varepsilon \right) $ in
gravity's~rainbow is same to general relativity, hence for a hypersurface,
we can use the Lanczos equation in the same form of the Eq. (\ref{sform}).
By using Eqs. (\ref{metric}), (\ref{kappa2}) and (\ref{kappa3}), the
extrinsic curvature tensor $\kappa _{ij\left( \text{rainbow}\right) }$ is
introduced in gravity's~rainbow by%
\begin{eqnarray}
\kappa _{\theta \left( \text{rainbow}\right) }^{\theta \pm } &=&\frac{\sqrt{%
h_{\varepsilon }^{2}e^{-2\lambda _{\pm }\left( r\right) }+\overset{.}{a}^{2}}%
}{a}, \\
&&  \notag \\
\kappa _{\tau \left( \text{rainbow}\right) }^{\tau \pm } &=&\frac{\phi _{\pm
}^{^{\prime }}\left( h_{\varepsilon }^{2}e^{-2\lambda _{\pm }\left( r\right)
}+\overset{.}{a}^{2}\right) +\ddot{a}+\overset{.}{a}^{2}\lambda _{\pm
}^{^{\prime }}}{\sqrt{h_{\varepsilon }^{2}e^{-2\lambda _{\pm }\left(
r\right) }+\overset{.}{a}^{2}}},
\end{eqnarray}%
and also%
\begin{eqnarray}
\sigma _{\left( \text{rainbow}\right) } &=&-\frac{\chi _{\theta }^{\theta }}{%
4\pi }  \notag \\
&&  \notag \\
&=&\frac{h_{\varepsilon }\left( \sqrt{e^{-2\lambda _{-}\left( r\right) }+%
\frac{\overset{.}{a}^{2}}{h_{\varepsilon }^{2}}}-\sqrt{e^{-2\lambda
_{+}\left( r\right) }+\frac{\overset{.}{a}^{2}}{h_{\varepsilon }^{2}}}%
\right) }{4\pi a},  \label{sigmaGsR} \\
&&  \notag \\
P_{\left( \text{rainbow}\right) } &=&\frac{\chi _{\tau }^{\tau }+\chi
_{\theta }^{\theta }}{8\pi }=  \notag \\
&&  \notag \\
&=&\frac{\left[ \frac{\left( 1+\phi ^{^{\prime }}a\right) \left(
h_{\varepsilon }^{2}e^{-2\lambda \left( r\right) }+\overset{.}{a}^{2}\right)
+a\ddot{a}+\lambda ^{^{\prime }}a\overset{.}{a}^{2}}{\sqrt{h_{\varepsilon
}^{2}e^{-2\lambda \left( r\right) }+\overset{.}{a}^{2}}}\right] ^{\pm }}{%
8\pi a}.  \label{PGsR}
\end{eqnarray}

The interior spacetime in the presence of dark energy should match the
exterior vacuum spacetime at a junction with $a$ radius. According to Eq. (%
\ref{metric}), we can write%
\begin{eqnarray}
e^{2\phi _{+}} &=&e^{-2\lambda _{+}}=1-\frac{2M}{r}, \\
&&  \notag \\
e^{-2\lambda _{-}} &=&1-\frac{2m\left( r\right) }{h_{\varepsilon }^{2}r}=1-%
\frac{2M_{eff}}{r},
\end{eqnarray}%
where $M_{eff}$ is the effective mass and it equals to $\frac{m\left(
r\right) }{h_{\varepsilon }^{2}}$ (i.e., $M_{eff}=\frac{m\left( r\right) }{%
h_{\varepsilon }^{2}}$). Thus, the exterior spacetime is followed%
\begin{equation}
ds_{+}^{2}=-\frac{1-\frac{2M}{r}}{l_{\varepsilon }^{2}}dt^{2}+\frac{dr^{2}}{%
h_{\varepsilon }^{2}\left( 1-\frac{2M}{r}\right) }+\frac{r_{+}^{2}\left(
d\theta ^{2}+\sin ^{2}\theta d\varphi ^{2}\right) }{h_{\varepsilon }^{2}},
\end{equation}%
where $M$ is total mass. In order to avoid the event horizon in the dark
energy star model, the junction radius places outside $a>2M$.

The surface energy density and the surface pressure, Eqs. (\ref{sigmaGsR})
and (\ref{PGsR}), can be obtained in term of $m(r)$, $M$ and $h_{\varepsilon
}$ as follows%
\begin{eqnarray}
\sigma _{\left( \text{rainbow}\right) } &=&\frac{h_{\varepsilon }\left(
\sqrt{1-\frac{2m\left( a\right) }{ah_{\varepsilon }^{2}}+\frac{\overset{.}{a}%
^{2}}{h_{\varepsilon }^{2}}}-\sqrt{1-\frac{2M}{a}+\frac{\overset{.}{a}^{2}}{%
h_{\varepsilon }^{2}}}\right) }{4\pi a},  \label{SigGsRII} \\
&&  \notag \\
P_{\left( \text{rainbow}\right) } &=&\frac{1}{8\pi a}\left[ \frac{%
h_{\varepsilon }^{2}\left( 1-\frac{M}{a}\right) +\overset{.}{a}^{2}+\ddot{a}%
}{\sqrt{h_{\varepsilon }^{2}\left( 1-\frac{2M}{a}\right) +\overset{.}{a}^{2}}%
}-\mathcal{P}\right] ,
\end{eqnarray}%
where $\mathcal{P}=\frac{\left( 1+\phi _{-}^{^{\prime }}a\right) \left(
h_{\varepsilon }^{2}\left( 1-\frac{2m\left( a\right) }{ah_{\varepsilon }^{2}}%
\right) +\overset{.}{a}^{2}\right) +a\ddot{a}+\frac{\left( am^{^{\prime
}}\left( a\right) -m\left( a\right) \right) \overset{.}{a}^{2}}{%
h_{\varepsilon }^{2}\left( a-\frac{2m\left( a\right) }{h_{\varepsilon }^{2}}%
\right) }}{\sqrt{h_{\varepsilon }^{2}\left( 1-\frac{2m\left( a\right) }{%
ah_{\varepsilon }^{2}}\right) +\overset{.}{a}^{2}}}$. From the above
equations, it can be seen that $\sigma $\ and $P$ depend on the rainbow
function $h_{\varepsilon }$ and are independent of $l_{\varepsilon }$.
According to the equation (\ref{6}), we can define the area of junction
sureface as $A_{\Sigma }=\frac{4\pi a^{2}}{h_{\varepsilon }^{2}}$, therefore
\textbf{the mass of a thin shell }$m_{s}=\sigma _{\left( \text{rainbow}%
\right) }A_{\Sigma }$\textbf{\ in gravity's rainbow is as follows}%
\begin{equation}
m_{s}=\frac{4\pi \sigma _{\left( \text{rainbow}\right) }a^{2}}{%
h_{\varepsilon }^{2}}.
\end{equation}%
By rewriting the Eq. (\ref{SigGsRII}) in terms of $M$, we can obtain the
total mass at a static radius $a_{0}$%
\begin{eqnarray}
M &=&M_{eff}\left( a_{0}\right)   \notag \\
&&  \notag \\
&&+m_{s}\left( a_{0}\right) h_{\varepsilon }\left( \sqrt{1-\frac{%
2M_{eff}\left( a_{0}\right) }{a_{0}}}-\frac{m_{s}\left( a_{0}\right)
h_{\varepsilon }}{2a_{0}}\right) .  \label{M2}
\end{eqnarray}

\begin{figure}[tbh]
\centering
\includegraphics[width=0.23\textwidth]{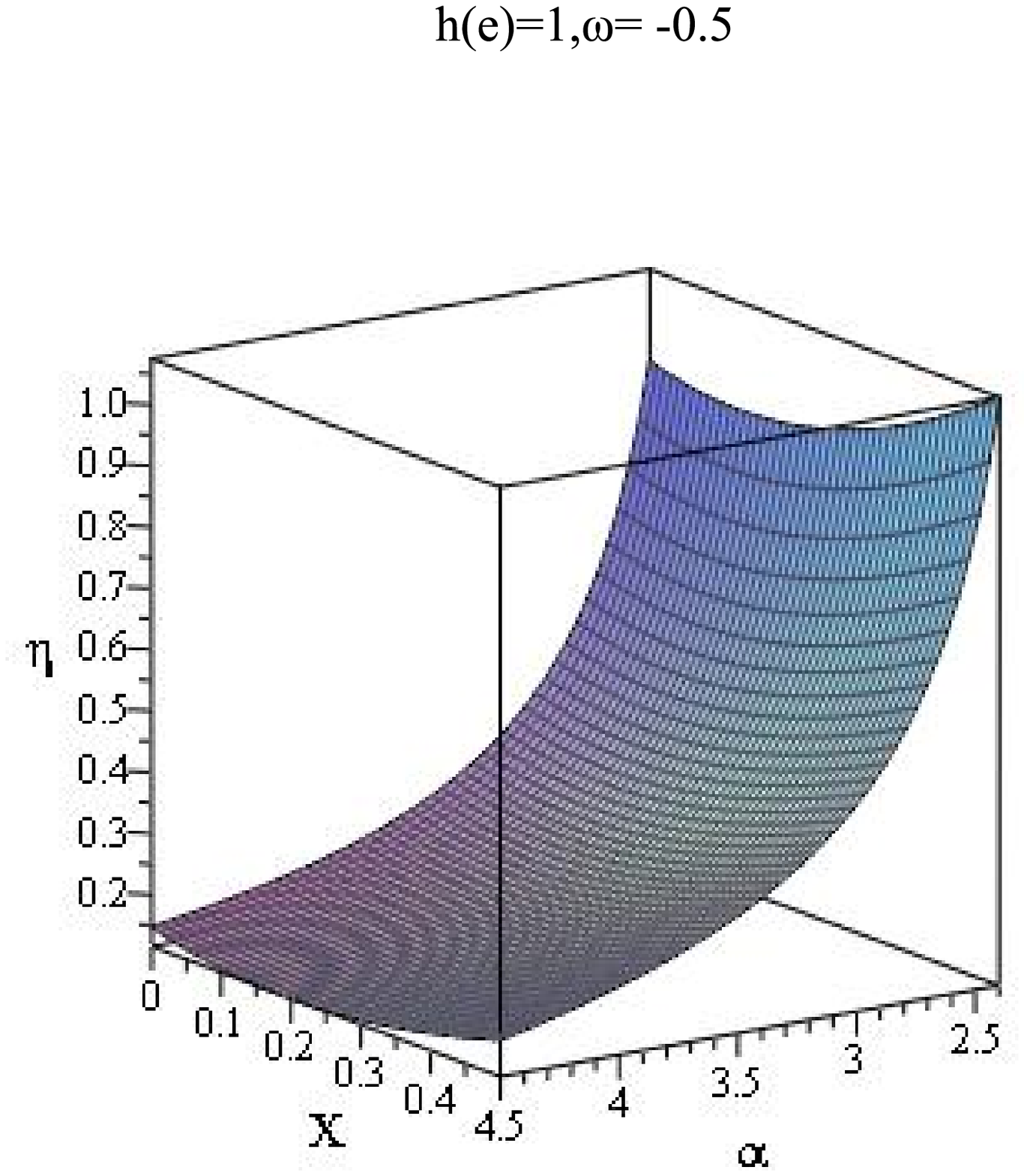} %
\includegraphics[width=0.23\textwidth]{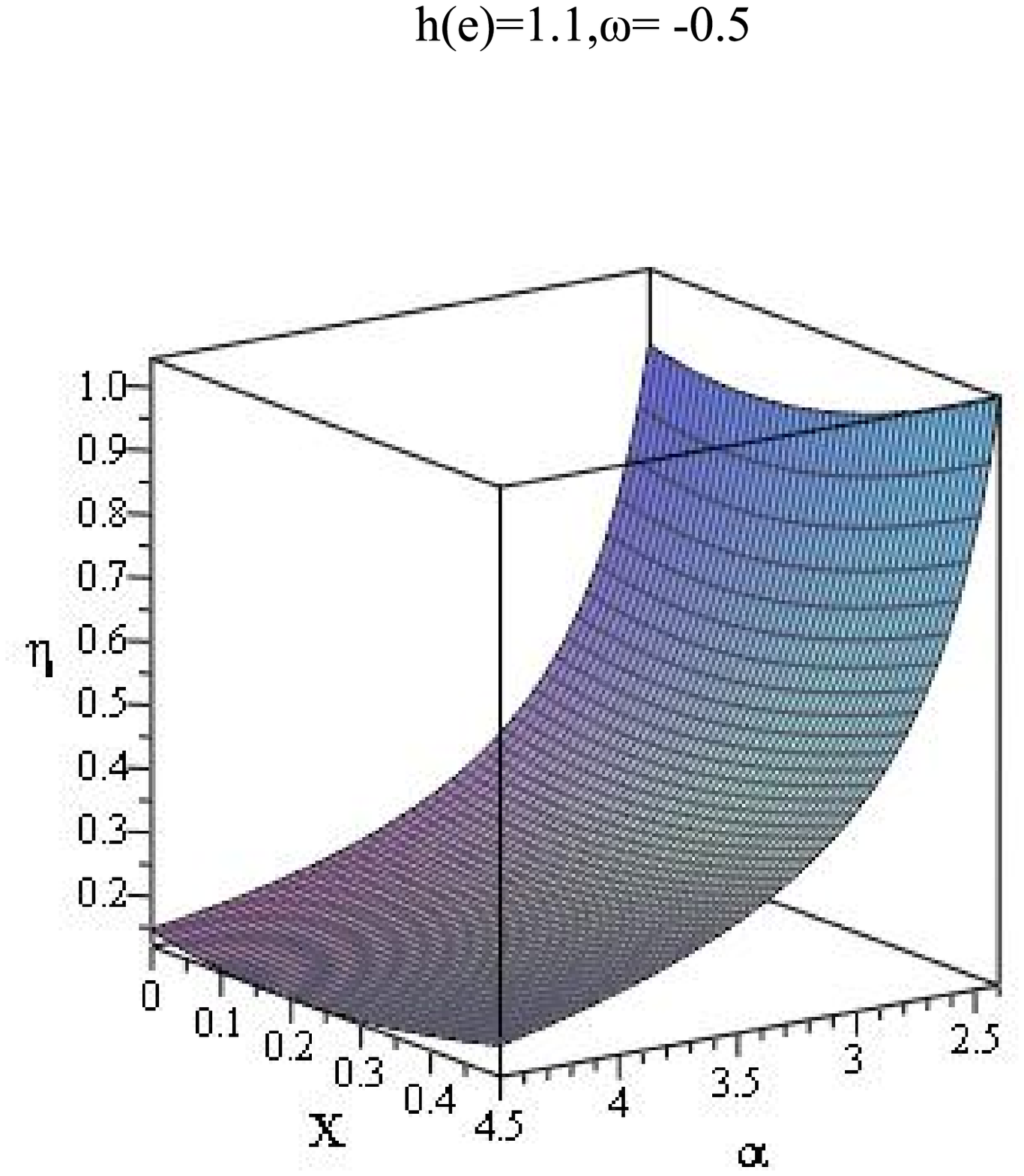} %
\includegraphics[width=0.23\textwidth]{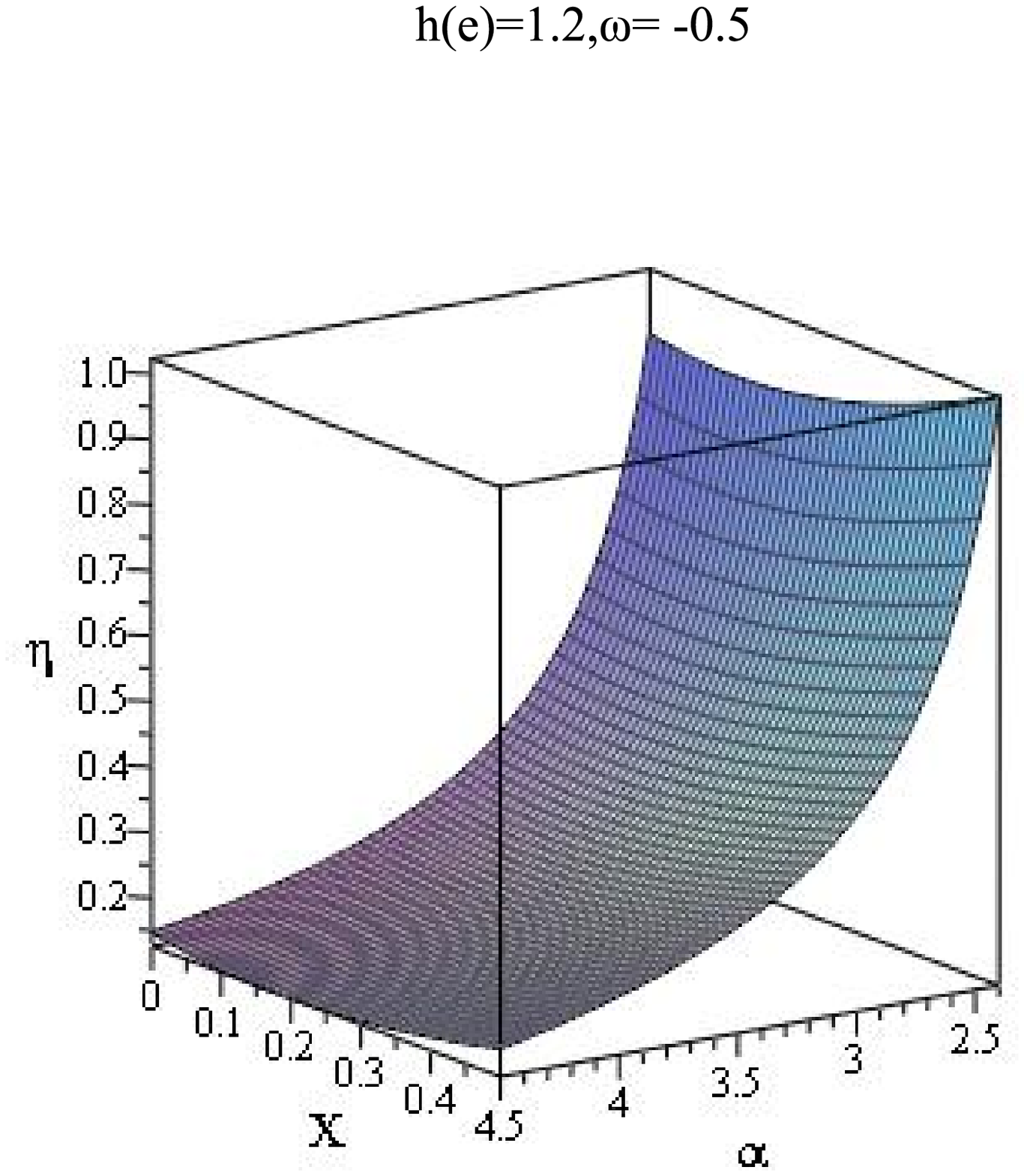} %
\includegraphics[width=0.23\textwidth]{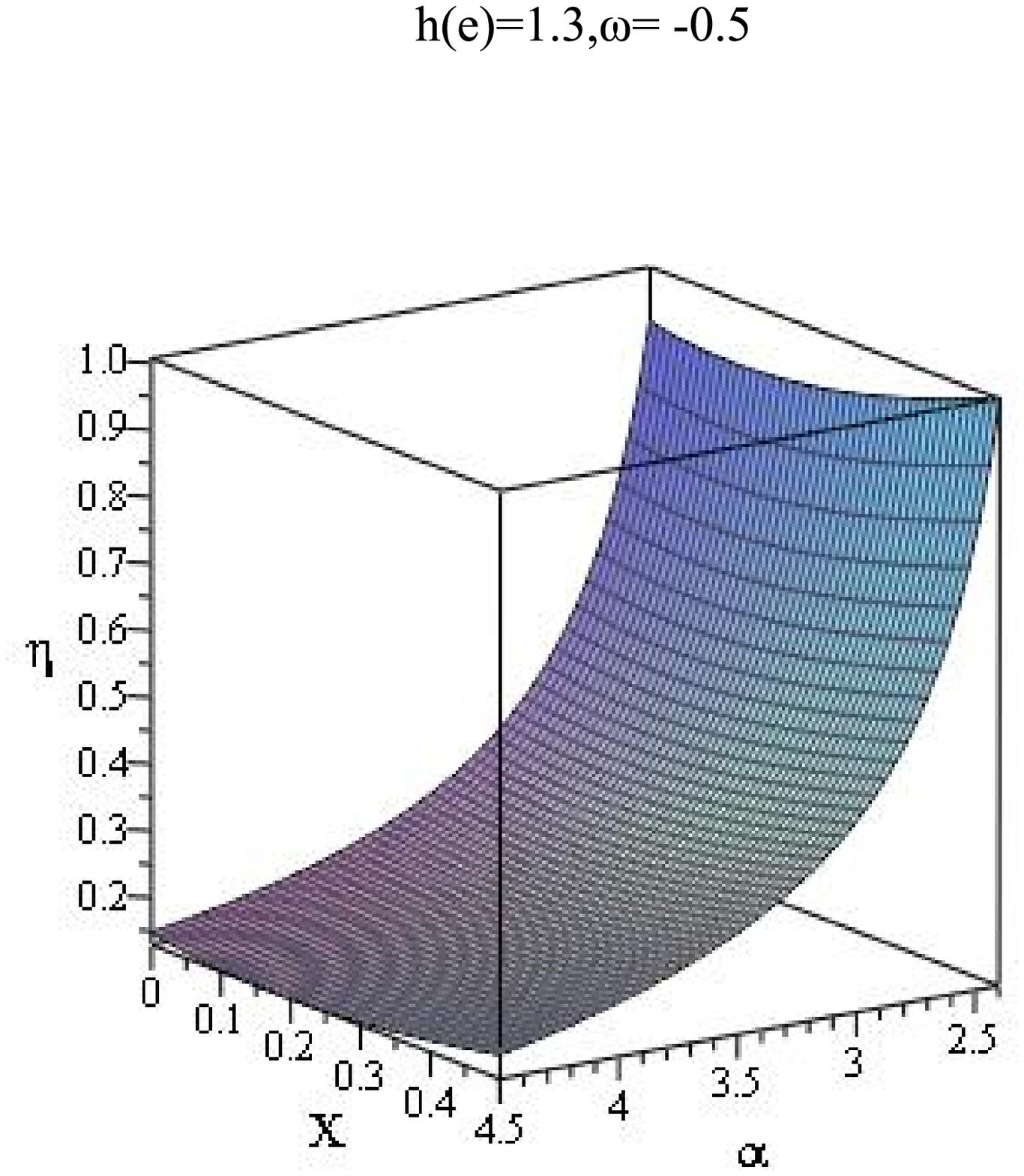} \newline
\caption{Three-dimensional diagrams of the stability regions. $X=\frac{m}{M}$
and $\protect\alpha=\frac{a}{M}$, are the dimensionless.}
\label{Fig6}
\end{figure}

\begin{figure}[tbh]
\centering
\includegraphics[width=0.22\textwidth]{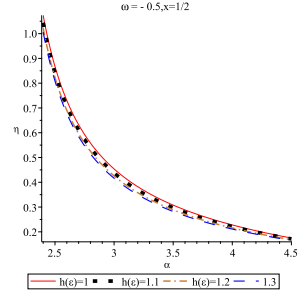} \newline
\caption{The stability region for DES versus $\protect\alpha=\frac{a}{M}$.}
\label{Fig7}
\end{figure}

As mentioned earlier, thin shell dynamical stability can be demonstrated
using parameter $\eta $. Fig. \ref{Fig6} shows the stability region for the
case $\omega =-0.5$ and different values of $h_{\varepsilon }$. It should be
noted that we can use $b_{0}=2m\left[ a^{3}\left( 1-4m/a\right) \right] ^{-1}
$ as an auxiliary tool \cite{Lobo2006}, thus $a>4m\left( a\right) $. On the
other hand, we assumed $a\gtrsim 2M$, so $0<\frac{m}{M}\lesssim \frac{1}{2}$%
. Note in these considerations, $h_{\varepsilon }$ is simplified. For $%
h_{\varepsilon }=1$, all equations yield to the usual form in general
relativity. It can be inferred from Fig. \ref{Fig7} that as the value of the
rainbow function $h_{\varepsilon }$ increases, the unstable regions near the
Schwarzschild radius move closer to stability. For values $h_{\varepsilon
}\geq 1.3$, the whole region is stable.

\section{ENERGY CONDITION}

\begin{figure*}[tbh]
\centering
\includegraphics[width=0.22\textwidth]{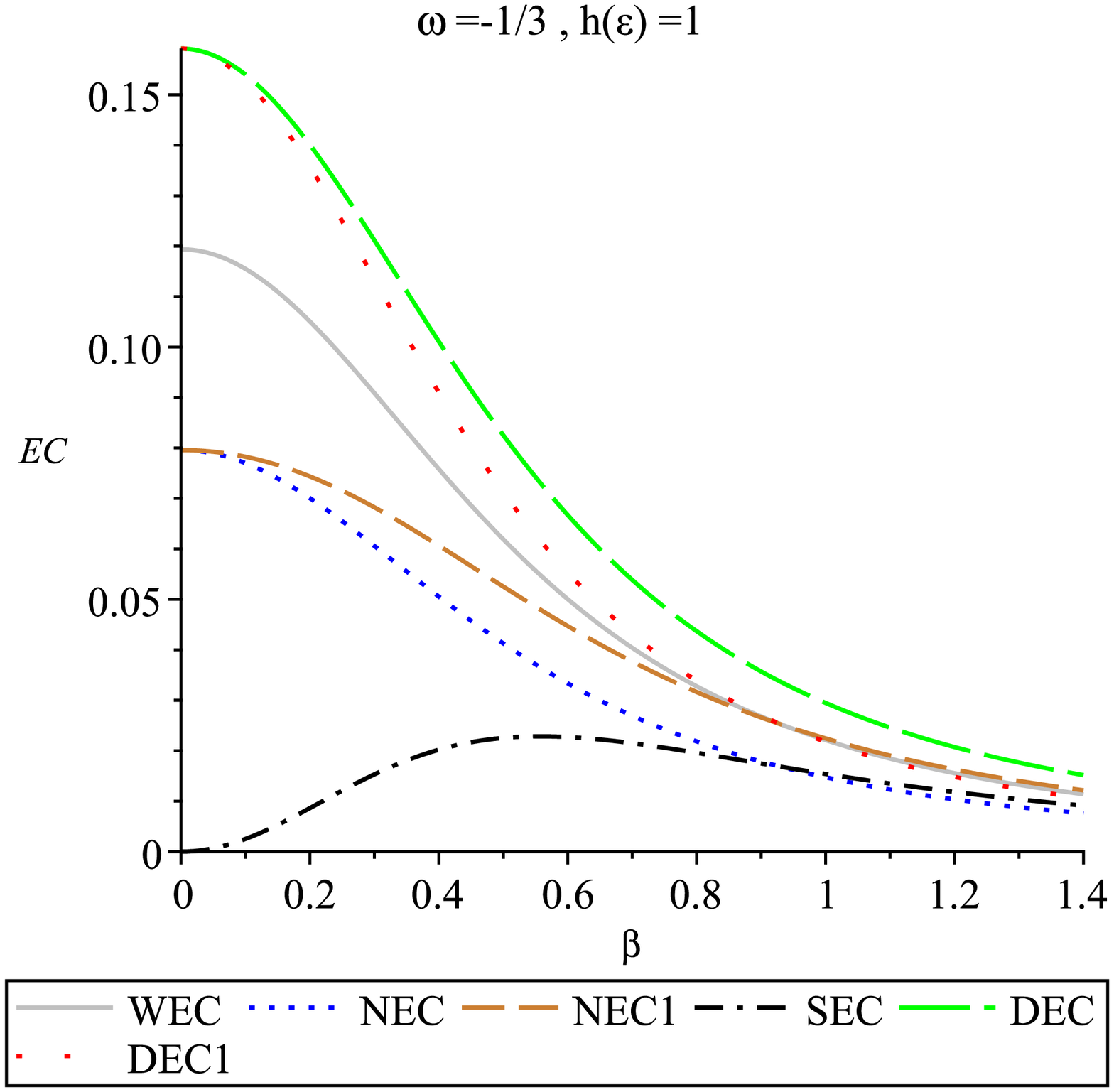} %
\includegraphics[width=0.22\textwidth]{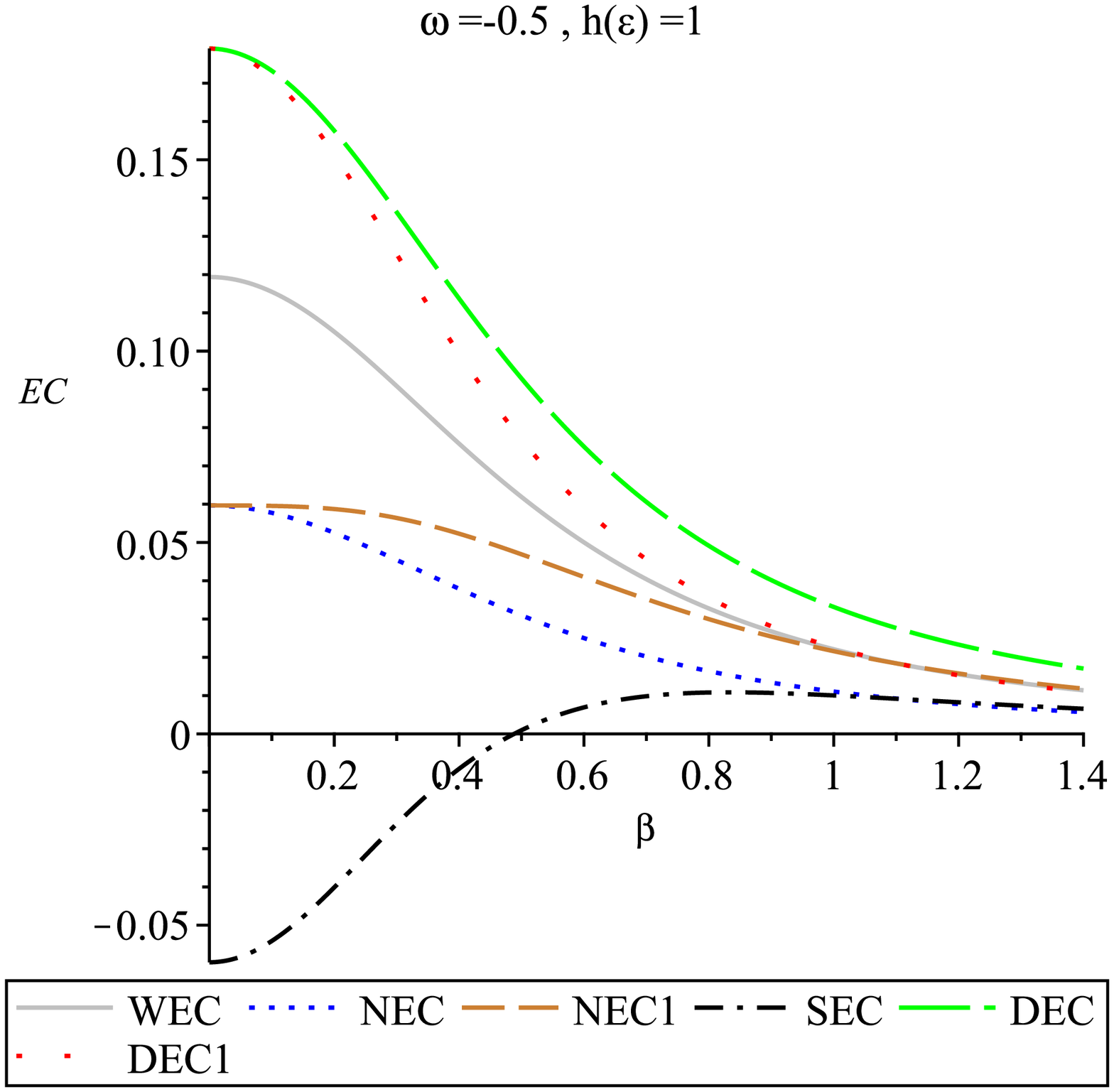} %
\includegraphics[width=0.22\textwidth]{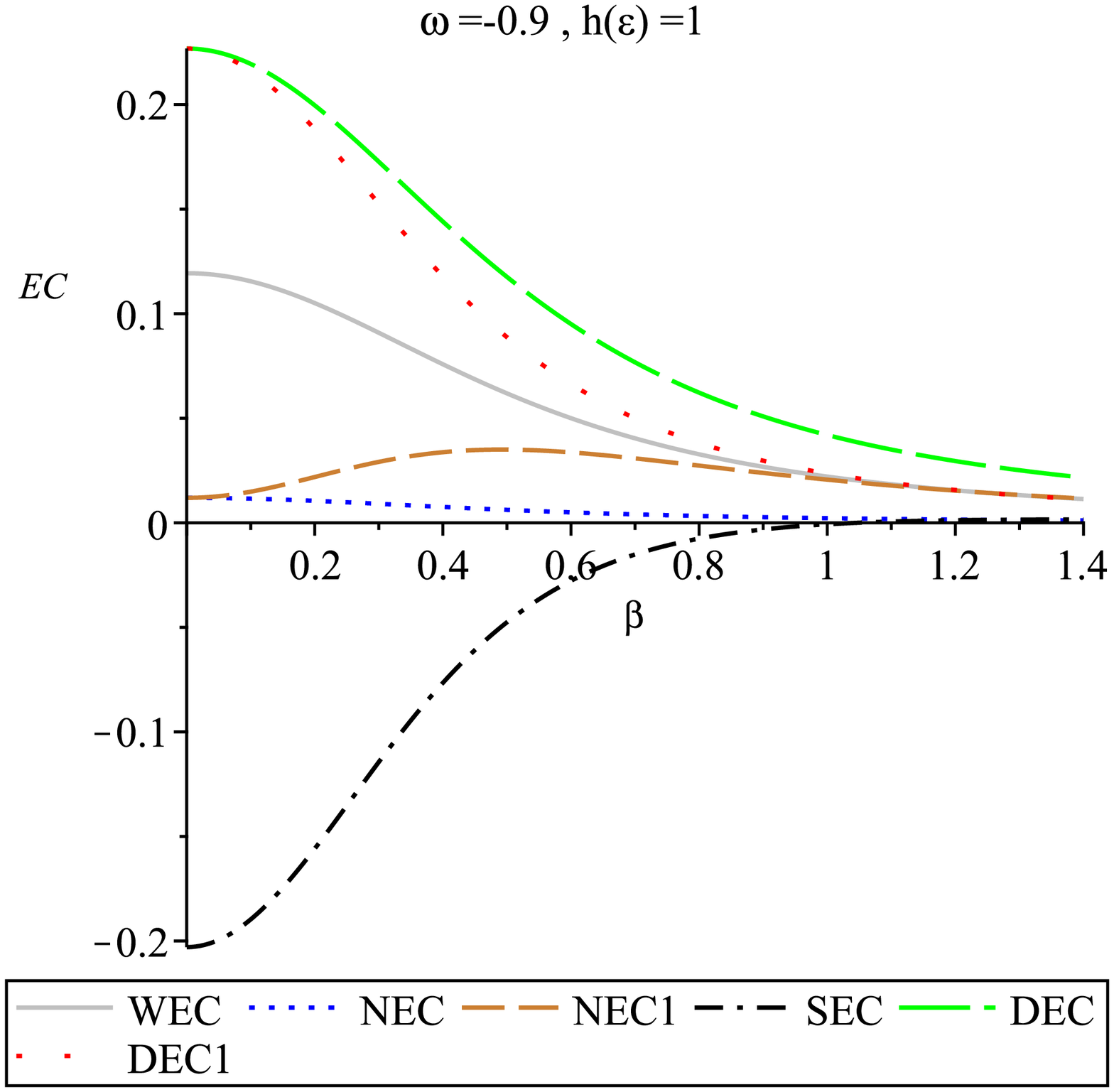} %
\includegraphics[width=0.22\textwidth]{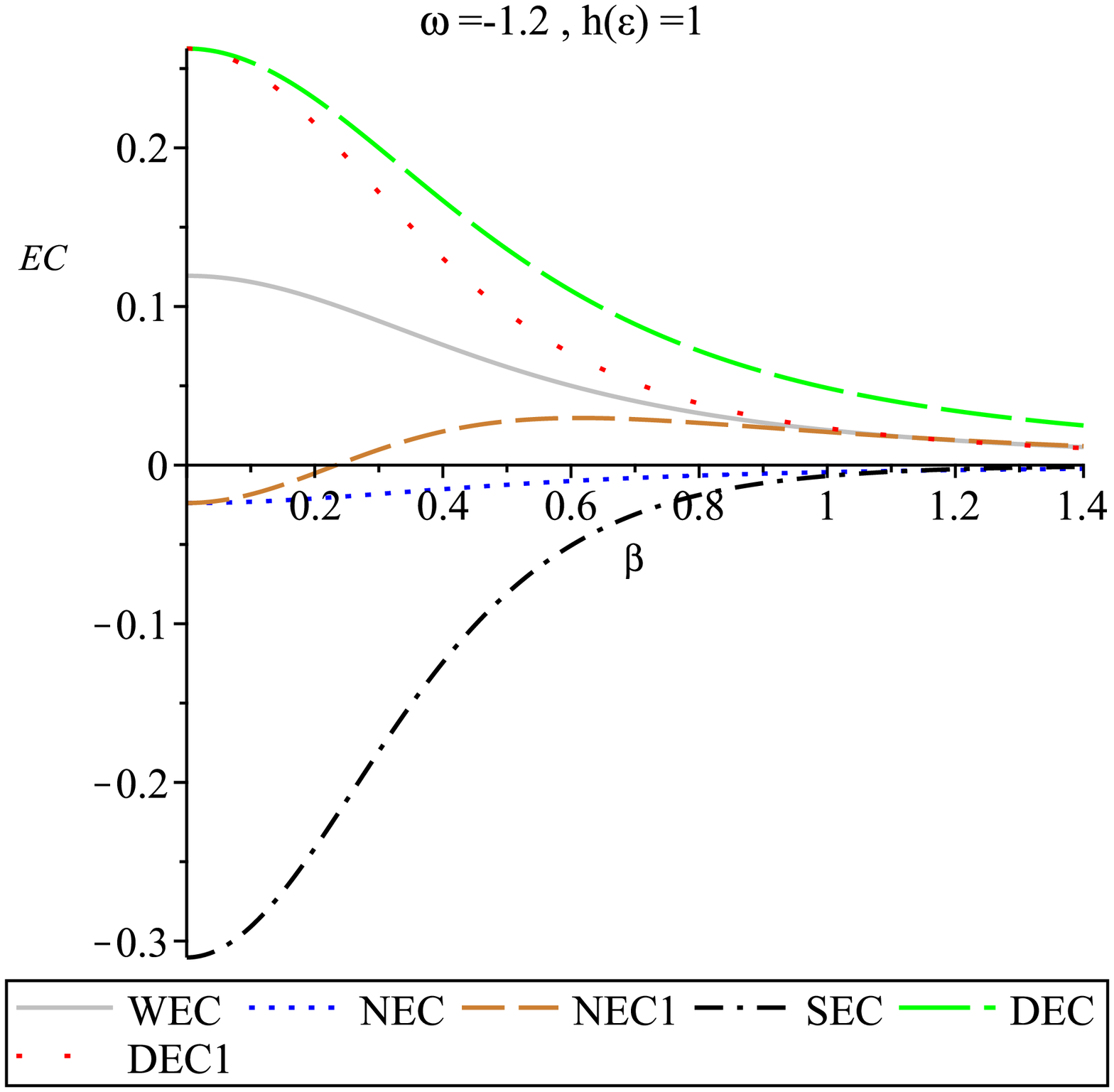} %
\includegraphics[width=0.22\textwidth]{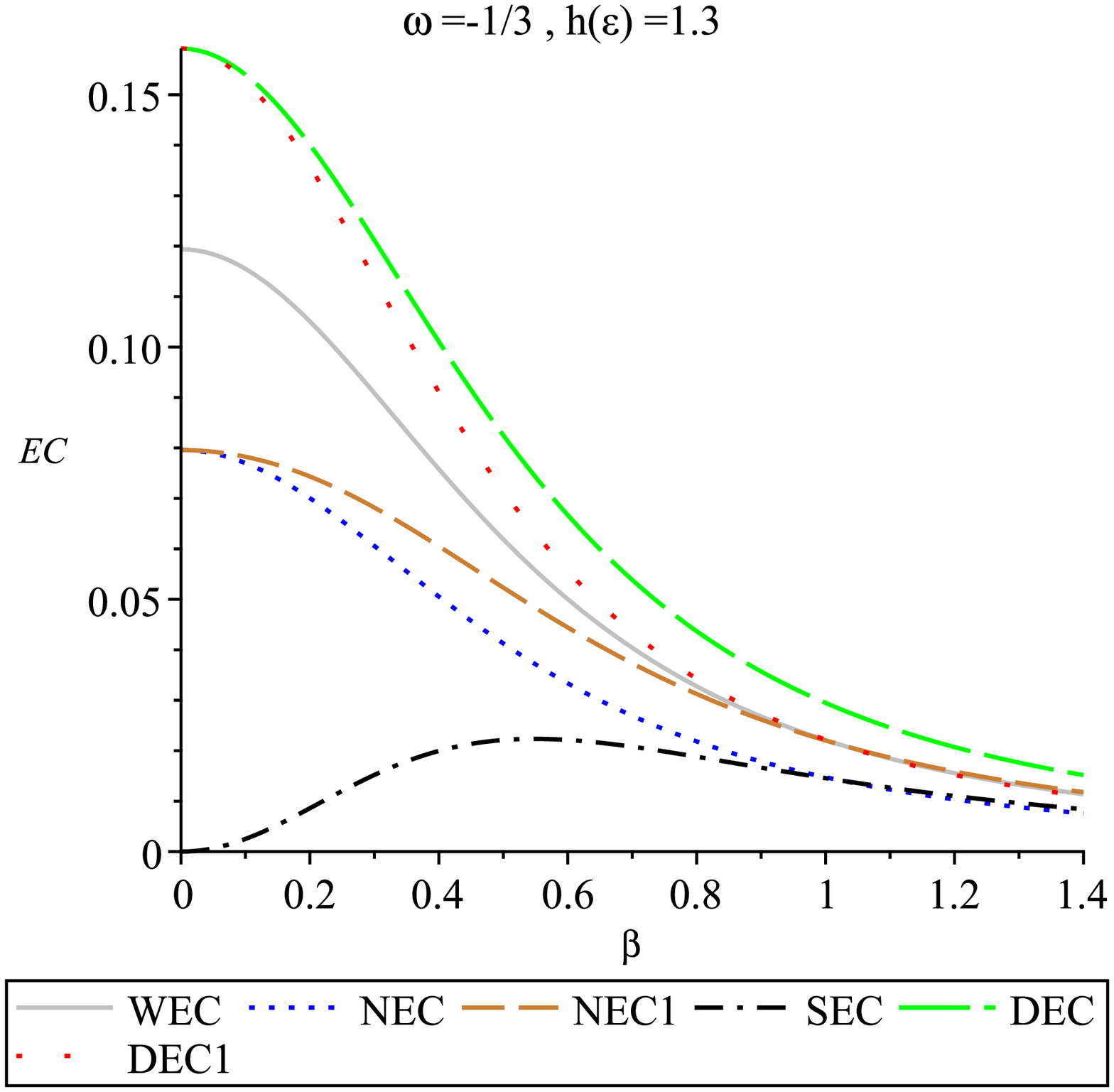} %
\includegraphics[width=0.22\textwidth]{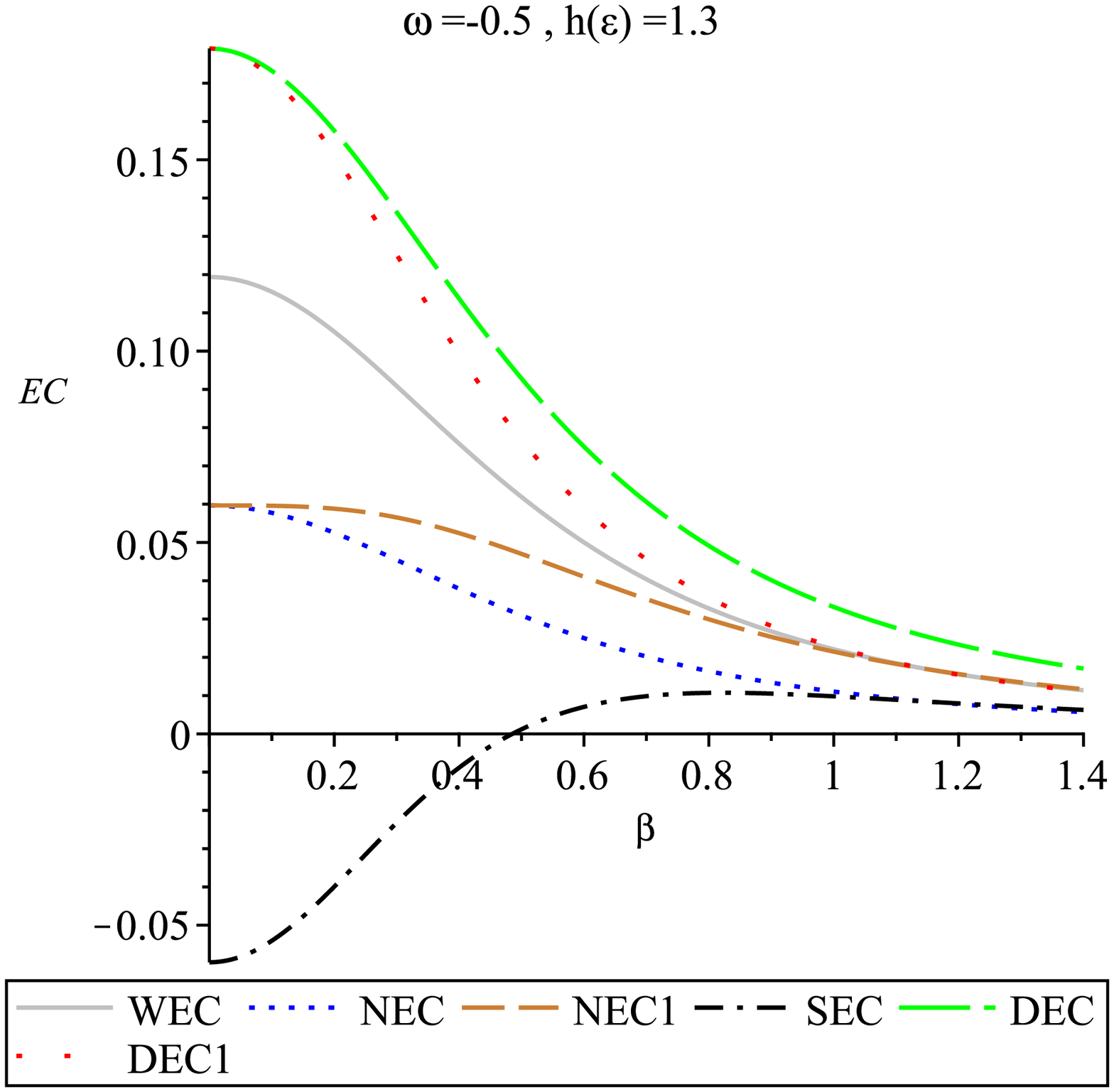} %
\includegraphics[width=0.22\textwidth]{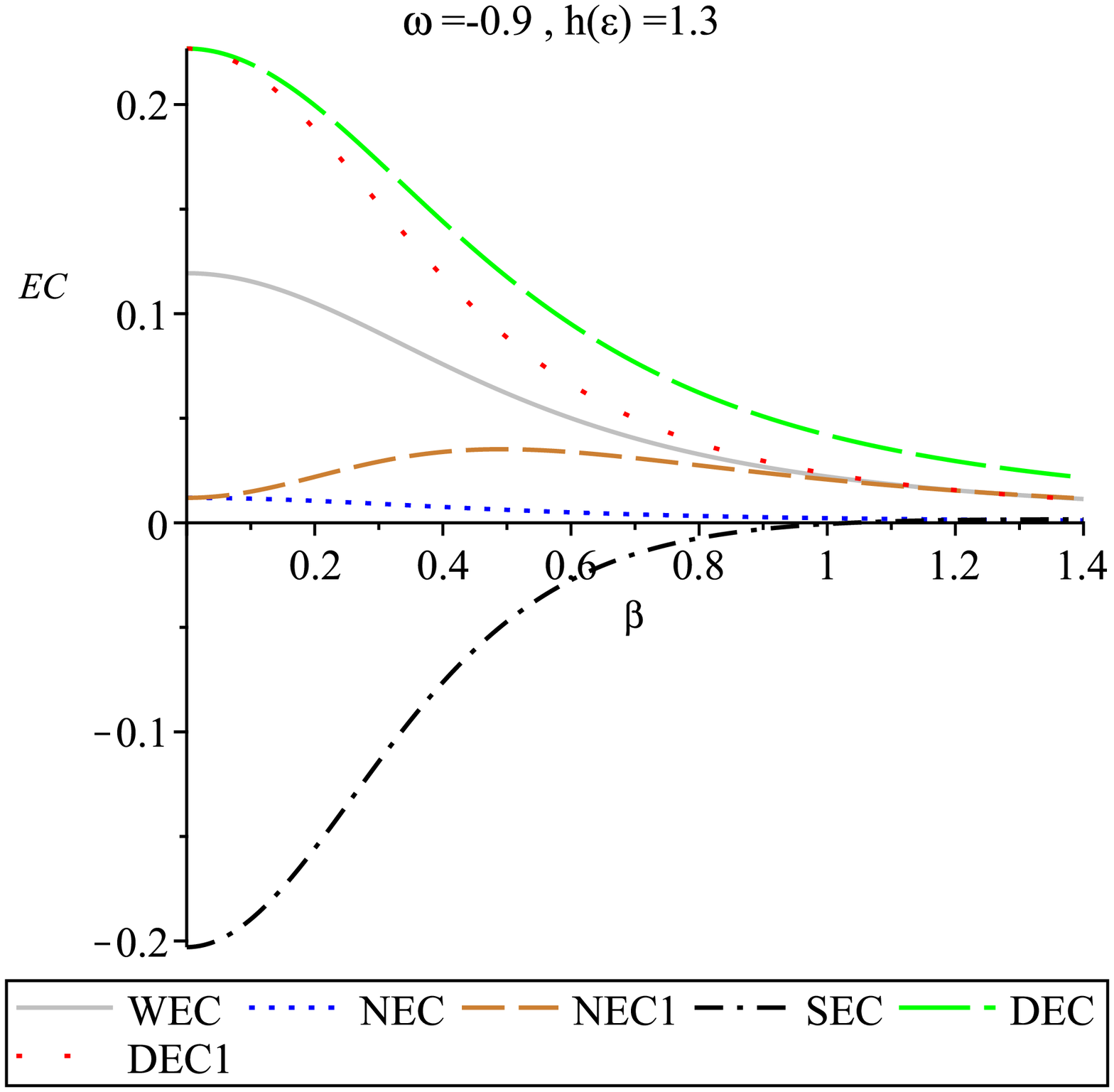} %
\includegraphics[width=0.22\textwidth]{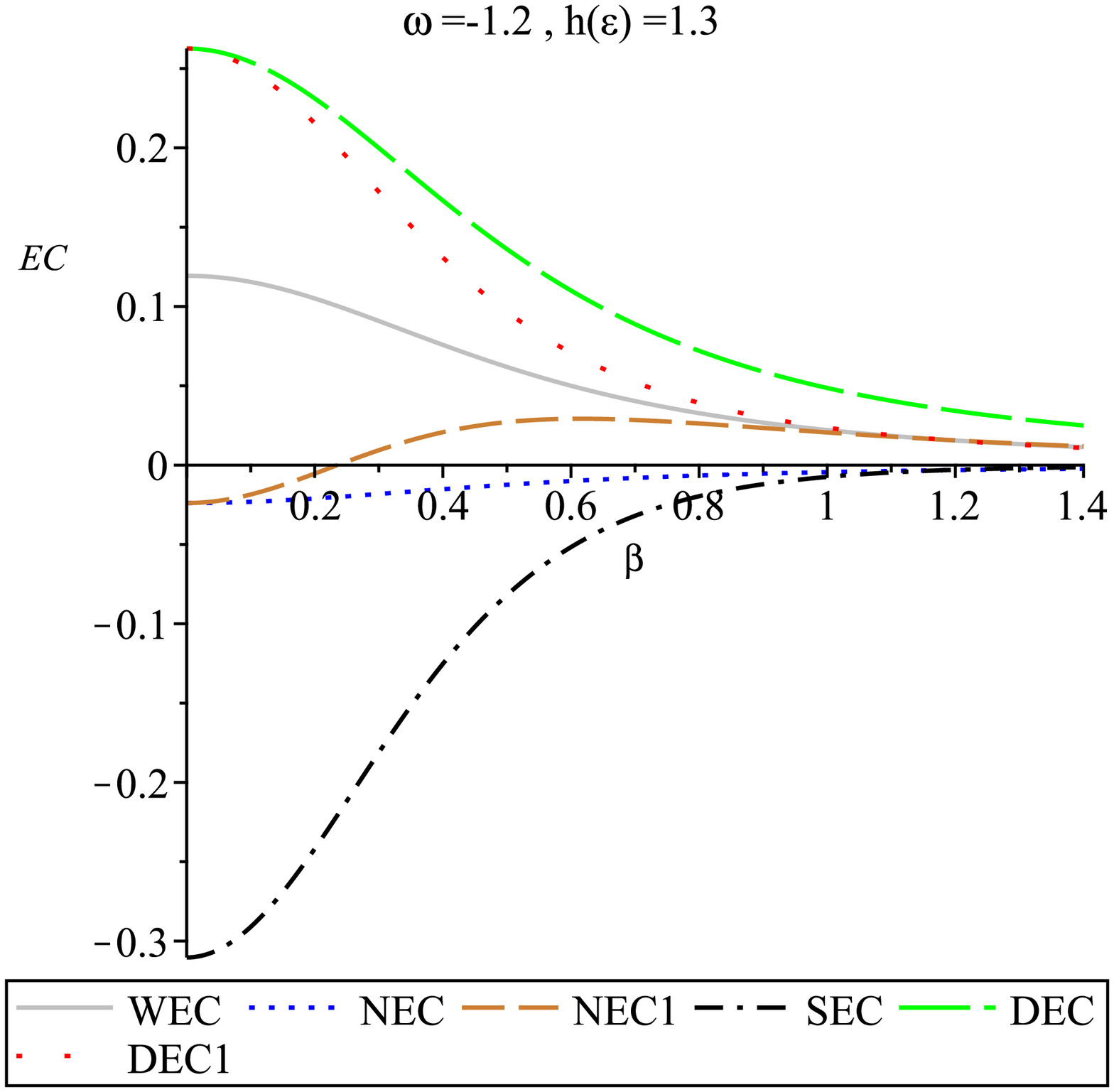} \newline
\caption{The energy conditions per $b_{0}$ in the interior of DES versus $%
\protect\beta =\protect\sqrt{b_{0}}r$. Note NEC1 and DEC1 refer to ($\protect%
\rho +p_{t}\geq 0$) and ($\protect\rho \geq \left\vert p_{t}\right\vert $),
respectively.}
\label{Fig8}
\end{figure*}

For interior region, the energy conditions are given by \cite%
{Leon1993,Visser1995} \newline
i) null energy condition (NEC): $\rho +p_{r}\geq 0$, and $\rho +p_{t}\geq 0$%
. \newline
ii) weak energy condition (WEC): $\rho \geq 0$, $\rho +p_{r}\geq 0$, and~$%
\rho +p_{t}\geq 0$. \newline
iii) strong energy condition (SEC): $\rho +p_{r}+2p_{t}\geq 0$, and~$\rho
+p_{t}\geq 0$. \newline
iv) dominant energy condition (DEC): $\rho \geq \left\vert p_{r}\right\vert $%
, and $\rho \geq \left\vert p_{t}\right\vert $. \newline
By placing Eqs. (\ref{rhoo}), (\ref{Pp}), and (\ref{Deltaa}) in above
definition for energy conditions, the interior energy conditions are
obtained
\begin{eqnarray}
\rho \left( r\right) +p_{r}\left( r\right)  &=&\frac{b_{0}\left(
2b_{0}r^{2}+3\right) \left( \omega +1\right) }{8\pi \left(
2b_{0}r^{2}+1\right) },  \notag \\
&&  \notag \\
\rho \left( r\right) -p_{r}\left( r\right)  &=&\frac{-b_{0}\left(
2b_{0}r^{2}+3\right) \left( \omega -1\right) }{8\pi \left(
2b_{0}r^{2}+1\right) },  \notag \\
&&  \notag \\
\rho \left( r\right) +p_{t}\left( r\right)  &=&\frac{\frac{b_{0}}{8\pi }%
\left[ \frac{8\mathcal{A}_{5}\left( b_{0}r^{2}+\frac{1}{2}\right)
h_{\varepsilon }^{2}}{b_{0}r^{2}}+\mathcal{A}_{6}\right] }{\left(
2b_{0}r^{2}+1\right) ^{3}\left[ \frac{\left( 2b_{0}r^{2}+1\right)
h_{\varepsilon }^{2}}{b_{0}r^{2}}-1\right] },  \notag \\
&&  \notag \\
\rho \left( r\right) -p_{t}\left( r\right)  &=&\frac{\frac{b_{0}}{8\pi }%
\left[ \frac{8\mathcal{A}_{7}\left( b_{0}r^{2}+\frac{1}{2}\right)
h_{\varepsilon }^{2}}{b_{0}r^{2}}-\mathcal{A}_{8}\right] }{\left(
2b_{0}r^{2}+1\right) ^{3}\left[ \frac{\left( 2b_{0}r^{2}+1\right)
h_{\varepsilon }^{2}}{b_{0}r^{2}}-1\right] },  \notag \\
&&  \notag \\
\rho \left( r\right) +p_{r}\left( r\right) +2p_{t}\left( r\right)  &=&\frac{%
\frac{b_{0}}{4\pi }\left[ \frac{4\mathcal{A}_{9}\left( b_{0}r^{2}+\frac{1}{2}%
\right) h_{\varepsilon }^{2}}{b_{0}r^{2}}+\mathcal{A}_{10}\right] }{\left(
2b_{0}r^{2}+1\right) ^{3}\left[ \frac{\left( 2b_{0}r^{2}+1\right)
h_{\varepsilon }^{2}}{b_{0}r^{2}}-1\right] },
\end{eqnarray}%
where $\mathcal{A}_{5}$, $\mathcal{A}_{6}$, $\mathcal{A}_{7}$, $\mathcal{A}%
_{8}$, $\mathcal{A}_{9}$ and $\mathcal{A}_{10}$ are%
\begin{eqnarray*}
\mathcal{A}_{5} &=&b_{0}^{2}r^{4}-\frac{b_{0}r^{2}\left( \omega -4\right) }{2%
}+\frac{3}{4}\left( \omega +1\right) , \\
&& \\
\mathcal{A}_{6} &=&b_{0}^{2}r^{4}\left( \omega -1\right) \left( \omega
+3\right)  \\
&&+3b_{0}r^{2}\left( \omega -\frac{2}{3}\right) \left( \omega +3\right) +%
\frac{9\left( \omega ^{2}-1\right) }{4}, \\
&& \\
\mathcal{A}_{7} &=&b_{0}^{2}r^{4}+\frac{b_{0}r^{2}\left( \omega +4\right) }{2%
}-\frac{3}{4}\left( \omega -1\right) , \\
&& \\
\mathcal{A}_{8} &=&b_{0}^{2}r^{4}\left( \omega ^{2}+2\omega -5\right)  \\
&&+3b_{0}r^{2}\left( \omega ^{2}+\frac{7\omega }{3}+\frac{10}{3}\right) +%
\frac{3\left( 3\omega ^{2}+5\right) }{4}, \\
&& \\
\mathcal{A}_{9} &=&b_{0}^{2}r^{4}\left( \omega +1\right) +b_{0}r^{2}\left(
\omega +2\right) +\frac{3\left( 3\omega +1\right) }{4}, \\
&& \\
\mathcal{A}_{10} &=&b_{0}^{2}r^{4}\left( \omega ^{2}-1\right)  \\
&&+3b_{0}r^{2}\left( \omega ^{2}+\omega -\frac{2}{3}\right) +\frac{3\left(
3\omega ^{2}-2\omega -1\right) }{4}.
\end{eqnarray*}%
The energy conditions are demonstrated in Fig. \ref{Fig8} for cases $%
h_{\varepsilon }=1$, and $h_{\varepsilon }=1.3$, respectively.

It can be seen in gravity's rainbow similar to general relativity, for $%
\omega =-\frac{1}{3}$, all energy conditions are obeyed by our calculations.
For $-\frac{1}{3}\leq \omega \leq -1$, all conditions are satisfied except
SEC. Violation of strong energy condition is a feature of dark energy. For $%
\omega <-1$, SEC and NES conditions are violated. Violation of null energy
condition is a feature of phantom energy.

\section{Conclusions}

In this study, by assuming the existence of an anisotropic distribution of
dark energy in the interior of spherically symmetric spacetime, we got a
modified TOV equation of anisotropic distribution in the gravity's rainbow.
In order to solve the energy-dependent field equations, we used the modified
Tolman-Matese-Whitman mass function. We considered the dark energy equations
of state to obtain the properties of dark energy stars. We showed that the
final solution is independent of the rainbow function $l_{\varepsilon }$,
and it only depends on $h_{\varepsilon }$. As the value of $h_{\varepsilon }$
increased, the gravity profile$\ g\left( r\right) $ became more constrained
thus the model is closer to the standard definition of a dark energy star,
and the anisotropy factor remain positive. As the rainbow function equals
one, this solution tends to the general relativity \cite{Lobo2006}. We also
investigated the dynamical stability of thin shell for the dark energy star
in this gravity by generalizing the Darmois-Israel formalism in the
gravity's rainbow. We showed that by increasing rainbow function $%
h_{\varepsilon }$, the unstable regions near the event horizon decrease. The
strong energy condition (SEC) is violated in the interior of the dark energy
star. It seems that employing the gravity's rainbow has the greatest effect
near where the phase transition zone is called. Where it is located near the
event horizon and high-energy particles decay there due to crossing the
critical surface \cite{Chapline,ChaplinearXiv}. The presence of modified
gravity with quantum gravity backgrounds will significantly help to study
the behavior of a dark energy star, especially near the critical region.

\begin{acknowledgements}
A. Bagheri Tudeshki and G. H. Bordbar wish to thank Shiraz University research council.
B. Eslam Panah thanks the University of Mazandaran. The University of Mazandaran has supported the work of B. Eslam Panah by
title "Evolution of the masses of celestial compact objects in various gravity".
\end{acknowledgements}

\end{document}